\documentclass[pra,twocolumn,superscriptaddress,showpacs,amsmath,amssymb]{revtex4-1}
\usepackage{epsfig}
\usepackage{subfigure}
\usepackage{amssymb,amsmath}
\usepackage{graphicx}
\usepackage{epstopdf}
\usepackage{epsfig}
\usepackage{color}
\usepackage{amsfonts}
\usepackage{amscd}
\usepackage{float}
\usepackage[colorlinks,urlcolor=blue,linkcolor=blue,citecolor=blue]{hyperref}
\usepackage{bbm}
\usepackage{mathrsfs}
\usepackage{mathtools}
\emergencystretch=\maxdimen
\DeclareMathSizes{10}{10}{7}{5}

\begin{document}
\title{Accessing the bath information in open quantum systems with the stochastic {\it c}-number Langevin equation method}
\date{\today}
\author{Zheng-Yang Zhou}
\affiliation{Theoretical Quantum Physics Laboratory, RIKEN Cluster for Pioneering Research, Wako-shi, Saitama 351-0198, Japan}

\author{Yun-An Yan}
\affiliation{
  School of Physics and Optoelectronic Engineering,
  Ludong University, Shandong 264025, China
}
\author{Stephen Hughes}
\affiliation{Department of Physics, Queen's University, Kingston, Ontario, Canada, K7L 3N6}
\author{J. Q. You}
\altaffiliation[jqyou@zju.edu.cn]{}
\affiliation{Department of Physics, Zhejiang University, Hangzhou 310027, China}
\author{Franco Nori}
\altaffiliation[fnori@riken.jp]{}
\affiliation{Theoretical Quantum Physics Laboratory, RIKEN Cluster for Pioneering Research, Wako-shi, Saitama 351-0198, Japan}
\affiliation{Physics Department, The University of Michigan, Ann Arbor, Michigan 48109-1040, USA}

\begin{abstract}
In traditional open quantum systems, the baths are usually traced out so that only the system information is left in the equations of motion. However, recent studies reveal that using only the system degrees of freedom can be insufficient. In this work, we develop a stochastic {\it c}-number Langevin equation method which can conveniently access the bath information. In our method, the studied quantities are the expectation values of operators which can contain both system operators and bath operators. The dynamics of the operators of interest is formally divided into separate system and bath parts, with auxiliary stochastic fields. After solving the independent stochastic dynamics of the system part and the bath part, we can recombine them by taking the average over these stochastic fields to obtain the desired quantities. Several applications of the theory are highlighted, including the pure dephasing model, the spin-boson model, and an optically excited quantum dot coupled to a bath of phonons.
\end{abstract}
%
%\pacs{??}
%
%\keywords{}
%
\maketitle

\section{Introduction}
Open quantum systems are usually divided into the most relevant parts (the system) and the secondary parts (the bath). Therefore one can use a partial trace to eliminate the bath. However, recent studies make the division between the system and the bath somewhat unclear. For example, the closed many-body systems do not have bath parts, but can still thermalize~\cite{mb2,mb3,mb4,mb5,mb6,mb7,mb8,mb9,mb10}. Large coupling strengths can make the correlation between the system and the bath important in thermodynamics~\cite{qtd1,qtd3,qtd4,qtd5,qtd6,qtd7,qtd8,qtd9}. In photosynthesis, the mixing of the system modes and the bath modes can suppress decoherence~\cite{qbio1,qbio2,qbio3,qbio4,photosyn1,photosyn2,photosyn3,photosyn4,photosyn5,photosyn5}. Moreover, when the bath is influenced by the pump exerted on the system, tracing out the bath is difficult and usually contains approximations~\cite{drib1,drib2,drib3}. Thus a method without performing a partial trace is very desirable.

To retain the potential information lost in tracing out the baths, one approach is to introduce generating functions~\cite{mebi1,mebi2,mebi3}. However, it is difficult to find the generating functions in general cases. The Heisenberg-Langevin approach~\cite{qle1,qle2,qle3,qle4,qle5,qle6} can be another good choice for such problems without apparent divisions between the systems and the baths. Although the Langevin equations consist of the system parts and the bath parts, no partial trace is included. Therefore, the accessible quantities are not restricted to the system parts~\cite{qleb1,qleb2,qleb3}. Nevertheless, the bath parts become the quantum noise terms in the equations, which makes the numerical simulations of the Langevin equations very difficult. Also, in some models, e.g., multi-level systems, the Langevin equations may contain the nonlinear time-non-local terms. Thus the use of the Heisenberg-Langevin approach in open quantum systems has limitations.

In the Schr$\rm \ddot{o}$dinger picture, nonlinear time-non-local terms can be avoided. Therefore, researchers have developed many successful methods, like the quantum state diffusion method~\cite{Diosi,qsd1,qsd2,qsd3,qsd4}, polaron transform methods~\cite{polatran1,polatran2,polatran3,polatran4,polatran5,polatran6,polatran7,polatran8,polatran9}, the hierarchy equation methods~\cite{hie1,hie2,hie3,hie4,hie5,hie5.5,hie6,hie7}, path integral methods~\cite{pathi}, and the stochastic Liouville equation methods~\cite{sfd0,sfd1,sfd2,sfd2.5,sfd3,sfd4,sfd5,oqs10}. These methods are suitable for different situations, but with various limitations. For example, the quantum state diffusion method can efficiently calculate large systems, but its application is difficult for many models. The polaron transform methods, which can provide compact master equations, usually include perturbations in either the transformations or the derivations of the master equations (which treats only certain parts of the bath coupling non-perturbatively). One can calculate the long-time results with hierarchy equation methods if the correlation functions of the baths are not too complicated. The path integral methods are powerful but complicated, and restricted when adding in other effects to the model. The stochastic methods are not restricted to any bath spectrum or temperature, but at the cost of poor long-time performance. However, most of the recent methods contain a partial trace.

In this article, we combine the Heisenberg-Langevin method and the stochastic Liouville equation to overcome some of the mentioned disadvantages of these recent methods. To avoid the nonlinear time-non-local terms in the equations, we separate the dynamics of the system and the bath with the Hubbard-Stratonovich transformation~\cite{hstrans} (by introducing auxiliary stochastic fields). Then the Langevin equations become linear and time local at the cost of containing classical noise terms. Moreover, we change the stochastic equations to {\it c}-number equations by taking expectation values. The final result can be obtained by taking the average over the noise terms. Our method is not based on the perturbation or the Markovian approximation, so it is, in principle, numerically exact. Similar to the stochastic Liouville equation, the stochastic {\it c}-number Langevin equations can deal with very complicated bath spectra at any temperature, but may have poor numerical convergence for long time simulations. In addition, our method also has the advantage of the Heisenberg-Langevin method , in that it can conveniently obtain the quantities of the bath.

Our paper is organized as follows. In Sec.~\ref{secsle}, we derive the stochastic {\it c}-number Langevin equations. We first separate the dynamics of the system and the bath by introducing the stochastic noise terms. Then, we solve the stochastic dynamics of the system part and the bath part separately. Finally, we combine the system part and the bath part by taking the average over the noise terms to obtain the desired results. In Sec.~\ref{example}, several numerical examples of two-level systems are calculated. The first example is the pure dephasing model which can be used to check the method. The second example, the spin-boson model, has been calculated with many numerical methods, but the bath quantities are seldom studied. Then a realistic and practical system, a quantum dot system coupled to phonons, is chosen as the third example. Section~\ref{conclusion} presents our conclusions. We also include four appendices. Appendix~\ref{AppA} provides the derivation of the stochastic bath evolution operator. In Appendix~\ref{AppB}, we describe how to generate the bath operators from the stochastic bath evolution operator. In Appendix~\ref{AppC}, we derive the expectation value of the stochastic identity operator of the bath. In Appendix~\ref{AppD}, our method is extended to the time-dependent Hamiltonian case.
\section{The stochastic {\it c}-number Langevin equations}
\label{secsle}
For the total system, we consider the Calderia-Leggett type model~\cite{c-l model}, with $H_{\rm tot}=H_{0}+H_{\rm I}$ (setting $\hbar=1$),
\begin{eqnarray}
H_{0}&=&H_{\rm sys}+H_{\rm b},\nonumber\\
H_{\rm b}&=&\sum_{k}\omega_ka^{\dag}_ka_k,\nonumber\\
H_{\rm I}&=&S\sum_{k}(g_k^*a_k^{\dag}+g_ka_k).\label{hamiltonian}
\end{eqnarray}
Here, $H_{0}$ is composed of the free Hamiltonian of the system, $H_{\rm sys}$, and the multi-mode bosonic bath $H_{\rm b}$. The coupling Hamiltonian $H_{\rm I}$ describes the coupling between the system and the bosonic bath. In the coupling Hamiltonian, $S$ is a Hermitian operator of the system, and $a_{ k}^\dagger$ ($a_{ k}$) is the bosonic creation (annihilation) operator of the $k$th-mode in the bath.

Solving such an open system with the Heisenberg-Langevin approach is usually difficult. Here, we take the standard generalized quantum Lanvegin equation as an example~\cite{osbook1}. Consider the Heisenberg equations of an arbitrary system operator $A_s(t)$, and the bath operator $a_k$, then
\begin{eqnarray}
\dot{A}_s(t)&=&i[H_{\rm sys},A_s(t)]+i[S,A_s(t)]\sum_{k}[g_k^*a_k^{\dag}(t)+g_ka_k(t)],\nonumber\\
\dot{a}_k(t)&=&-ig_kS(t)-i\omega_ka_k.
\end{eqnarray}
The bath equations can be formally solved,
\begin{eqnarray}
a_k=e^{-i\omega_kt}a_k(0)-ig_k\int_0^t\!\!dse^{-i\omega_k(t-s)}S(s).
\end{eqnarray}
Substituting this formal solution for the
bath operators in the equation of the system operator $A_s(t)$, the Lanvegin equation of $A_s(t)$ can be derived:
\begin{eqnarray}
\dot{A}_s(t)&=&i[H_{\rm sys}(t),A_s(t)]+i[S(t),A_s(t)]\hat{\xi}(t)\nonumber\\
            &&+[S(t),A_s(t)]\int_0^t\!\!ds[\alpha(t-s)-\alpha^*(t-s)]S(s),\nonumber\\\label{qlefms}
\end{eqnarray}
where,
\begin{eqnarray}
\alpha(t-s)\equiv\sum_kg_kg_k^*\exp(-i\omega_k(t-s)),
\end{eqnarray}
is the zero-temperature correlation function, and,
\begin{eqnarray}
\hat{\xi}(t)\equiv\sum_k\left[g_ka_k(0)\exp(-i\omega_kt)+g_k^*a_k^{\dag}(0)\exp(i\omega_kt)\right],\nonumber\\
\end{eqnarray}
is the quantum noise. In this article, noise terms with hat correspond to quantum noise terms of the bath, and noise terms without hat are classical noise terms terms from the auxiliary field. In general, the commutator $[S,A_s(t)]$ is some operator $B_s(t)$. Thus  the last term in Eq.~(\ref{qlefms}) has the form $\int_0^tds[\alpha(t-s)-\alpha^*(t-s)]B_s(t)S(s)$, which is both nonlinear and time non-local. Such a term makes it difficult to apply the Heisenberg-Lanvegin approach to general cases. In addition, the quantum noise further increases the difficulty of solving these equations.
\subsection{Separating the system and the bath with stochastic noise terms}
To avoid dealing with the nonlinear terms directly, we formally separate the dynamics of the system and the bath~\cite{sfd1,sfd2,sfd2.5,sfd3,sfd4,sfd5}. The basic idea is similar to generating indirect interactions with intermediate fields~\cite{idcoupling1,idcoupling2,idcoupling3,idcoupling4}. Instead of the original system-bath coupling, we consider the equivalent coupling induced by auxiliary fields with trivial dynamics. This produces stochastic, but uncoupled system dynamics and bath dynamics. By averaging over the noise terms, which resembles tracing out the intermediate field, we can recover the original system-bath coupling.

An operator at time $t$ in the Heisenberg picture $B(t)$ is related to its initial value $B(0)$ by
\begin{eqnarray}
B(t)=U^{\dag}(t)B(0)U(t),
\end{eqnarray}
with the evolution operator $U(t)\equiv \exp(-iH_{\rm tot}t)$. It is difficult to obtain $U(t)$ for general cases. However, by applying the Hubbard-Stratonovich transformation, the evolution operator $U(t)$ can be divided into two parts,
\begin{widetext}
\begin{eqnarray}
U(t)&=&\int D[z_{1,\tau}]D[z_{2,\tau}]\frac{1}{2\pi}\exp\!\left({\int_0^t\!\!d\tau(-\frac{1}{2}z_{1,\tau}^2-\frac{1}{2}z_{2,\tau}^2)}U_{\rm sys}(t;z_{1,\tau},z_{2,\tau})U_{\rm b}(t;z_{1,\tau},z_{2,\tau})\right)\nonumber\\
    &\equiv&\mathcal{M}_z\left\{U_{\rm sys}(t;z_{1,\tau},z_{2,\tau})U_{\rm b}(t;z_{1,\tau},z_{2,\tau})\right\},\nonumber
\end{eqnarray}
where,
\begin{eqnarray}
U_{\rm sys}(t;z_{1,\tau},z_{2,\tau})&=&T_{+}\exp\!\left({-i\int_0^t\!\!d\tau\left[H_{\rm sys}
     +\frac{1}{\sqrt{2}}S(z_{1,\tau}+iz_{2,\tau})\right]}\right),\nonumber
\end{eqnarray}
and,
\begin{eqnarray}
U_{\rm b}(t;z_{1,\tau},z_{2,\tau})&=&T_{+}\exp\!\left({-i\int_0^t\!\!d\tau\left[\frac{1}{\sqrt{2}}\sum_{k}(g_ka_k+g_k^*a_k^\dag)(iz_{1,\tau}+z_{2,\tau})+\sum_{k}\omega_ka^{\dag}_ka_k\right]}\right).\nonumber\\\label{sfd}
\end{eqnarray}
\end{widetext}
Here, $U_{\rm sys}(t;z_{1,\tau},z_{2,\tau})$ is the stochastic evolution operator of the system, and $U_{\rm b}(t;z_{1,\tau},z_{2,\tau})$ is the stochastic evolution operator of the bath; $T_+$ is the time ordered operator, and the average over the noise terms is denoted by $\mathcal{M}_z\{\}$. We first assume $B(0)=B_{\rm sys}(0)\otimes B_{\rm b}(0)$ so that the dynamics of the total system can be divided into two parts. Thus:
\begin{eqnarray}
B(t)&=&\mathcal{M}_z\left\{B_{\rm sys}(t;z)\otimes B_{\rm b}(t;z)\right\},\nonumber\\
B_{\rm sys}(t;z)&=&U^{\dag}_{\rm sys}(t;z_{1,\tau},z_{2,\tau})B_{\rm sys}(0)U_{\rm sys}(t;z_{3,\tau},z_{4,\tau}),\nonumber\\
B_{\rm b}(t;z)&=&U^{\dag}_{\rm b}(t;z_{1,\tau},z_{2,\tau})B_{\rm b}(0)U_{\rm b}(t;z_{3,\tau},z_{4,\tau}).\nonumber\\\label{deoperator}
\end{eqnarray}
In more general cases, with $B(0)=\sum_iB_{i,\rm sys}(0)\otimes B_{i,\rm b}(0)$, we can deal with these terms one by one.

In the master equation method, the bath part is usually traced out to obtain the reduced density matrix $\rho_{\rm sys}(t)$. However, some information of the bath is lost in this procedure. To avoid such a disadvantage, we take the expectation value of the operator, instead of tracing out the bath:
\begin{eqnarray}
\langle B(t)\rangle=\rm{Tr}\left\{B(t)\rho_{\rm tot}(0)\right\}.
\end{eqnarray}
Now we assume that the system and the bath are factorized at the initial time $\rho_{\rm tot}(0)=\rho_{\rm sys}(0)\otimes\rho_{\rm b}(0)$. Then, the expectation value can also be divided into the system part and the bath part,
\begin{eqnarray}
\langle B(t)\rangle&=&\rm{Tr}\left\{\mathcal{M}_z\left\{B_{\rm sys}(t;z)\otimes B_{\rm b}(t;z)\right\}\rho_{\rm sys}(0)\otimes\rho_{\rm b}(0)\right\}\nonumber\\
                   &=&\mathcal{M}_z\left\{\langle B_{\rm sys}(t;z)\rangle\langle B_{\rm b}(t;z)\rangle\right\},\label{stoexp}
\end{eqnarray}
with,
\begin{eqnarray}
\langle B_{\rm sys}(t;z)\rangle\equiv\rm{Tr}\left\{B_{\rm sys}(t;z)\rho_{\rm sys}(0)\right\},
\end{eqnarray}
and,
\begin{eqnarray}
\langle B_{\rm b}(t;z)\rangle\equiv\rm{Tr}\left\{B_{\rm b}(t;z)\rho_{\rm b}(0)\right\}.
\end{eqnarray}
According to Eq.~(\ref{stoexp}), we can separately calculate the stochastic expectation value of the system part $\langle B_{\rm sys}(t;z)\rangle$ and the bath part $\langle B_{\rm b}(t;z)\rangle$. Then, the expectation value of the concerned operator $\langle B(t)\rangle$ can be obtained by taking the noise average of the stochastic ones. Note that these noise terms come from the auxiliary fields instead of the bath. The quantum noise induced by the bath is included in $\langle B_{\rm b}(t;z)\rangle$.
\subsection{The stochastic dynamics of the system part}
We first consider a system of finite dimension. In such a case, a set of basis operators $Y_{l}$ can be found to represent the $B_{\rm sys}(0;z)$,
\begin{eqnarray}
B_{\rm sys}(0;z)&=&\sum_lb_lY_l.
\end{eqnarray}
The stochastic system operator at time $t$ can also be expressed in the same way
\begin{eqnarray}
B_{\rm sys}(t;z)&=&U^{\dag}_{\rm sys}(t;z_{1,\tau},z_{2,\tau})\sum_lb_lY_lU_{\rm sys}(t;z_{3,\tau},z_{4,\tau})\nonumber\\
                &=&\sum_lb_lY_l(t;z),\nonumber\\
                {\rm with},~~~~~~~~~~&&\nonumber\\
Y_l(t;z)&\equiv&U^{\dag}_{\rm sys}(t;z_{1,\tau},z_{2,\tau})Y_lU_{\rm sys}(t;z_{3,\tau},z_{4,\tau}).\nonumber\\\label{sobasis}
\end{eqnarray}
The expectation value of the system part can then be expressed as
\begin{equation}
\langle B_{\rm sys}(t;z)\rangle=\sum_lb_l\langle Y_l(t;z)\rangle.\label{esobasis}
\end{equation}
According to Eq.~(\ref{esobasis}), once all the stochastic expectation values of the basis $Y_l(t;z)$ are known, the stochastic expectation value of the system part can be easily calculated. The equation for $Y_{l}(t;z)$ can be obtained by directly taking the time derivative of it, so that
\begin{eqnarray}
\frac{\partial}{\partial t}Y_{l}(t;z)&=&\frac{\partial}{\partial t}\left\{U^{\dag}_{\rm sys}(t;z_{1,\tau},z_{2,\tau})Y_{l}U_{\rm sys}(t;z_{3,\tau},z_{4,\tau})\right\},\nonumber\\
                                           &=&\left[\frac{\partial}{\partial t}U^{\dag}_{\rm sys}(t;z_{1,\tau},z_{2,\tau})\right]Y_lU_{\rm sys}(t;z_{3,\tau},z_{4,\tau})\nonumber\\
                                           &&+U^{\dag}_{\rm sys}(t;z_{1,\tau},z_{2,\tau})Y_{l}\left[\frac{\partial}{\partial t}U_{\rm sys}(t;z_{3,\tau},z_{4,\tau})\right].\nonumber\\\label{efsb}
\end{eqnarray}
According to the form of the stochastic evolution operator in Eq.~(\ref{sfd}), Eq.~(\ref{efsb}) can be written as
\begin{eqnarray}
\frac{\partial}{\partial t}Y_l(t;z)&=&U^{\dag}_{\rm sys}(t;z_{1,\tau},z_{2,\tau})D(0;z)U_{\rm sys}(t;z_{3,\tau},z_{4,\tau}),\nonumber
\end{eqnarray}
where,
\begin{eqnarray}
~D(0;z)&=&i[H_{\rm sys}
     +\frac{1}{\sqrt{2}}S_{\rm sys}(z_{1,t}-iz_{2,t})]Y_l\nonumber\\
        &&-iY_{l}[H_{\rm sys}
     +\frac{1}{\sqrt{2}}S_{\rm sys}(z_{3,t}+iz_{4,t})].\nonumber\\\label{ssle1}
\end{eqnarray}
If we define two complex noise terms:
\begin{eqnarray}
x_{1,t}&=&\frac{1}{2}(z_{1,t}-iz_{2,t}+z_{3,t}+iz_{4,t}),\nonumber
\end{eqnarray}
and,
\begin{eqnarray}
x_{2,t}&=&\frac{1}{2}(z_{1,t}-iz_{2,t}-z_{3,t}-iz_{4,t}).\nonumber\\
\end{eqnarray}
We can express $D(0;z)$ in Eq.(\ref{ssle1}) in a compact form,
\begin{eqnarray}
D(0;x)&=&i[H_{\rm sys},Y_l]+i\frac{1}{\sqrt{2}}x_{1,t}[S_{\rm sys},Y_l]\nonumber\\
       &&+i\frac{1}{\sqrt{2}}x_{2,t}\{S_{\rm sys},Y_l\}.\nonumber\\
\end{eqnarray}
The commutators and anti-commutators can also be expressed by the basis operators $Y_m$,
\begin{eqnarray}
[H_{\rm sys},Y_l]&=&\sum_{m}\mathcal{H}_{lm}Y_m,\nonumber\\
{[}S_{\rm sys},Y_l{]}&=&\sum_{m}\mathcal{S}^c_{lm}Y_{m},\nonumber\\
\{S_{\rm sys},Y_l\}&=&\sum_{m}\mathcal{S}^a_{lm}Y_{m}.\label{cmatrix}
\end{eqnarray}
When the dimension of the system Hilbert space is infinite, our method can also be applied if a set of basis operators which satisfies Eq.~(\ref{cmatrix}) can be found. With Eq.~(\ref{ssle1}) and Eq.~(\ref{cmatrix}), we can obtain the following equation for $Y_l(t;x)$
\begin{eqnarray}
\frac{\partial}{\partial t}Y_{l}(t;x)=i\sum_m\left(\mathcal{H}_{lm}+\frac{x_{1,t}}{\sqrt{2}}\mathcal{S}^c_{lm}+\frac{x_{2,t}}{\sqrt{2}}\mathcal{S}^a_{lm}\right)Y_{m}(t;x).\nonumber\\
\end{eqnarray}
The equation for the expectation values is straightforward to obtain,
\begin{eqnarray}
&&\frac{\partial}{\partial t}\langle Y_{l}(t;x)\rangle\nonumber\\
&=&i\sum_m\left(\mathcal{H}_{lm}+\frac{x_{1,t}}{\sqrt{2}}\mathcal{S}^c_{lm}+\frac{x_{2,t}}{\sqrt{2}}\mathcal{S}^a_{lm}\right)\langle{Y}_{m}(t;x)\rangle.\nonumber\\\label{ssle2}
\end{eqnarray}
Equation~(\ref{ssle2}) can also be written in a vector form if we define the stochastic expectation value vector as
$$\mathcal{Y}(t,x)\equiv(\langle Y_{1}(t;x)\rangle,\ldots,\langle Y_{n}(t;x)\rangle)^T,$$
where $n$ is the number of the basis operators of the system and $T$ means transpose. The vector form of the equation reads
\begin{equation}
\frac{\partial}{\partial t}\mathcal{Y}(t,x)=i\left(\mathcal{H}+\frac{x_{1,t}}{\sqrt{2}}\mathcal{S}^c+\frac{x_{2,t}}{\sqrt{2}}\mathcal{S}^a\right)\mathcal{Y}(t,x).\label{ssle3}
\end{equation}
The elements of the matrices $\mathcal{H}$, $\mathcal{S}^c$ and $\mathcal{S}^a$ are, respectively, the terms $\mathcal{H}_{lm}$, $\mathcal{S}^c_{lm}$, and $\mathcal{S}^a_{lm}$ in Eq.~(\ref{ssle2}). With Eq.~(\ref{esobasis}) and Eq.~(\ref{ssle3}), the stochastic expectation value of the system part can be calculated. Note that the choice of the basis operators is arbitrary. For example, we can take $B_{\rm sys}(t;z)$ as the first one of them, and find enough operators to satisfy Eq.~(\ref{cmatrix}).
\subsection{The Stochastic Dynamics of the Bath Part}
Next, we come to the stochastic evolution operator for the bath $U_{\rm b}(t;z_{3,\tau},z_{4,\tau})$. In the following, for simplicity, the stochastic evolution operators for the bath will be written as $U_{\rm Ib}(t;z)$ and $U_{\rm Ib}^{\dag}(t;z)$. Note that the noise terms are different in $U_{\rm Ib}(t;z)$ and $U_{\rm Ib}^{\dag}(t;z)$. We first change to the interaction picture,
\begin{eqnarray}
U_{\rm Ib}(t;z)=e^{iH_{\rm b}t}U_{\rm b}(t;z).\\\nonumber
\end{eqnarray}
Then the bath part of the stochastic operator becomes $B_{\rm b}(t;z)=U_{\rm Ib}^{\dag}(t;z)B_{\rm Ib}(t)U_{\rm Ib}(t;z)$, where $B_{\rm Ib}(t)\equiv\exp({iH_{\rm b}t})B_{\rm b}(0)\exp({-iH_{\rm b}t})$
is the initial value of the bath part of the stochastic operator in interaction picture. The equation for ${U}_{\rm Ib}(t;z)$ is
\begin{eqnarray}
\frac{\partial}{\partial t}{U}_{\rm Ib}(t;z)&=&\frac{\partial}{\partial t}\left[e^{iH_{\rm b}t}U_{\rm b}(t;z)\right]\nonumber\\
                                              &=&\frac{z_{3,\tau}-iz_{4,\tau}}{\sqrt{2}}\hat{\xi}(t){U}_{\rm Ib}(t;z).\label{efsbo}
\end{eqnarray}
Here the $\hat{\xi}(t)$ is the quantum noise term which appears in the quantum Langevin equation. Equation~(\ref{efsbo}) can be solved by the Magnus expansion~\cite{mge,mge2} (also see Appendix A),
\begin{widetext}
\begin{eqnarray}
{U}_{Ib}(t;z)&=&\exp\!\left(-i\frac{1}{\sqrt{2}}\int_0^t\!\! ds\,(iz_{3,s}+z_{4,s})\sum_kg_{k}^*e^{i\omega_ks}a^\dag_{kb}\right)\exp\!\left(-i\frac{1}{\sqrt{2}}\int_0^t\!\! ds\,(iz_{3,s}+z_{4,s})\sum_kg_{k}e^{-i\omega_ks}a_{kb}\right)\times\nonumber\\
&&\exp\!\left(-\int_0^t\!\!ds_1\int_0^{s_1}\!\!ds_2\,\frac{1}{2}(iz_{3,s_1}+z_{4,s_1})(iz_{3,s_2}+z_{4,s_2})\,\alpha(s_1-s_2)\right).
\label{sbo2}
\end{eqnarray}
\end{widetext}
Now we consider the stochastic expectation value of the bath part $\langle B_{\rm b}(t;z)\rangle$. We express the bath part of the operator with a Taylor expansion,
\begin{equation}
B_{\rm b}(0)=\Pi_k\sum_{m_k,n_k}C_{k,m_k,n_k}a_{\rm kb}^{n}a_{\rm kb}^{\dag{m}}.\label{pebo}
\end{equation}
The stochastic bath part of operator at time $t$ is,
\begin{equation}
B_{\rm b}(t;z)=\Pi_k\sum_{m_k,n_k}C_{k,m_k,n_k}U_{\rm b}^{\dag}(t;z)a_{\rm kb}^{n}a_{\rm kb}^{\dag m}U_{\rm b}(t;z).\label{pebot}
\end{equation}
Conceptually Eq.~(\ref{pebot}) can be directly calculated from Eq.~(\ref{sbo2}). However, the bath operators can also be generated approximately from the stochastic evolution operator when the time is not too short (see Appendix B). In this way,
\begin{eqnarray}
a^{\dag}_{\rm kb}U_{\rm b}(t;z)&=&\mathcal{A}_k(t;z_{3,\tau},z_{4,\tau})U_{\rm b}(t;z),\nonumber\\
a_{\rm kb}U^{\dag}_{\rm b}(t;z)&=&\mathcal{A}^*_k(t;z_{1,\tau},z_{2,\tau})U_{\rm b}^{\dag}(t;z),\nonumber\\
\end{eqnarray}
where $\mathcal{A}_k(t;z_{3,\tau},z_{4,\tau})$ has the following form,
\begin{eqnarray}
&&\mathcal{A}_k(t;z_{3,\tau},z_{4,\tau})\nonumber\\
&=&\frac{i\sqrt{2}e^{i\omega_kt}}{g_k^*}\int_0^t\!\!ds_1\!\left[e^{-i\omega_ks_1}\frac{\delta}{\delta z_{4,s_1}}\right.\nonumber\\
&&+\frac{1}{2}\!\int_0^{s_1}\!\!ds_2(iz_{3,s_2}+z_{4,s_2})e^{-i\omega_ks_1}\alpha(s_1-s_2)\nonumber\\
&&\left.+\frac{1}{2}\!\int_{s_1}^t\!\!ds_2(iz_{3,s_2}+z_{4,s_2})e^{-i\omega_ks_1}\alpha^*(s_1-s_2)\right]\!.\label{gsooad}
\end{eqnarray}
The term $\alpha(s_1-s_2)\equiv\sum_{k}g_kg_k^*\exp(-i\omega_k(s_1-s_2))$ in Eq.~(\ref{gsooad}) is the zero-temperature correlation function mentioned in Eq.~(\ref{qlefms}). With Eq.~(\ref{gsooad}),  Eq.~(\ref{pebot}) can be converted to
\begin{eqnarray}
B_{\rm b}(t;z)&=&\Pi_k\sum_{m_k,n_k}C_{k,m_k,n_k}\times\nonumber\\
&&[\mathcal{A}^*_k(t;z)]^mU_{\rm b}^{\dag}(t;z)[\mathcal{A}_k(t;z)]^nU_{\rm b}(t;z)\nonumber\\
              &=&\Pi_k\sum_{m_k,n_k}C_{k,m_k,n_k}[\mathcal{A}_k(t;z)]^n[\mathcal{A}^*_k(t;z)]^m I_{\rm b}(t;z).\nonumber\\\label{pebot2}
\end{eqnarray}
Here, $I_{\rm b}(t;z)\equiv U_{\rm b}^{\dag}(t;z)U_{\rm b}(t;z)$ is the bath part of the stochastic identity operator. Note that the stochastic identity operator is no longer an identity operator. Also, we abbreviate the $\mathcal{A}_k(t;z_{3,\tau},z_{4,\tau})$ [$\mathcal{A}^*_k(t;z_{1,\tau},z_{2,\tau})$] to $\mathcal{A}_k(t;z)$ [$\mathcal{A}^*_k(t;z)$]. Since the stochastic bath operators can be generated from the bath part of the stochastic identity operator, the next step is to obtain the stochastic expectation value of $I_{\rm b}(t;z)$. If the initial state of the bath is the thermal state, then $\langle I_{\rm b}(t;z)\rangle$ can be exactly calculated~\cite{sfd5}(also see Appendix C):
\begin{eqnarray}
\langle I_b(t;z)\rangle&=&\exp\!\left(\int_0^t\!\!ds\,g(s,z)\,x^*_{1,s_s}\right),\nonumber\\
g(s,z)&=&\int_0^s\!\!ds_1\left[(z_{3,s_1}-iz_{4,s_1})\,\alpha_T(s-s_1)\right.\nonumber\\
        &&\left.+(z_{1,s_1}+iz_{2,s_1})\,\alpha^*_T(s-s_1)\right],\nonumber
\end{eqnarray}
{\rm with},
\begin{eqnarray}
\alpha_T(t-s)&=&\sum_kg_kg_k^*\coth\!\left(\frac{\beta\omega_k}{2}\right)\cos(\omega_k(s_1-s_2))\nonumber\\
        &&-i\sum_kg_kg_k^*\sin(\omega_k(s_1-s_2)).\nonumber\\\label{seobu}
\end{eqnarray}
The $\alpha_T(t-s)$ term in Eq.~(\ref{seobu}) is the finite temperature correlation function of the bath. The effects of the bath temperature $T$ is described by the parameter $\beta\equiv1/({k_{\rm B}T})$. With the value of $\langle I_b(t;z)\rangle$, the stochastic expectation value of the bath part $\langle B_{\rm b}(t;z)\rangle$ can be calculated from:
\begin{eqnarray}
&&\langle B_{\rm b}(t;z)\rangle\nonumber\\
&=&\Pi_k\sum_{m_k,n_k}C_{k,m_k,n_k}[\mathcal{A}_k(t;z)]^n[\mathcal{A}^*_k(t;z)]^m \langle I_{\rm b}(t;z)\rangle.\nonumber\\
\end{eqnarray}
Using the exponential property of $\langle I_{\rm b}(t;z)\rangle$, we can assume that $\langle B_{\rm b}(t;z)\rangle$ has the following form,
\begin{eqnarray}
\langle B_{\rm b}(t;z)\rangle=f(t,x_3)\langle I_{\rm b}(t;z)\rangle.
\end{eqnarray}
The $f(t,x_3)$ term is some stochastic function, where $x_3$ represents one or more noise terms. The property of $x_3$ depends on the form of $B_{\rm b}(t;z)$. If we are only interested in the system operators, then the bath part is just $B_{\rm b}(t;z)=I_{\rm b}(t;z)$.
\subsection{Noise average of the stochastic operators}
To obtain the expectation value of the operator $B(t)$, we need to eliminate the auxiliary stochastic fields by taking a noise average over the product of the system part and the bath part,
\begin{eqnarray}
\langle B(t)\rangle&=&\mathcal{M}_z\left\{\langle B_{\rm sys}(t;z)\rangle\langle B_{\rm b}(t;z)\rangle\right\}\nonumber\\
                   &=&\mathcal{M}_z\left\{\sum_lb_l\langle{Y}_l(t;x)\rangle f(t,x_3)\langle{I}(t;z)\rangle\right\}.\nonumber\\\label{saoo1}
\end{eqnarray}
Direct calculation of Eq.~(\ref{saoo1}) is numerically inefficient. We can introduce a transformation of the noise terms to absorb $\langle I(t;z)\rangle$ into the relevant distribution function~\cite{sfd2},
\begin{eqnarray}
z_{1,t}'&=&z_{1,t}-\frac{1}{2}g(t,z),\nonumber\\
z_{2,t}'&=&z_{2,t}-\frac{i}{2}g(t,z),\nonumber\\
z_{3,t}'&=&z_{3,t}-\frac{1}{2}g(t,z),\nonumber\\
z_{4,t}'&=&z_{4,t}+\frac{i}{2}g(t,z).\nonumber\\\label{tron}
\end{eqnarray}
Notice that this transformation only changes the noise terms in the system part. With the transformation in Eq.~(\ref{tron}), Eq.~(\ref{ssle3}) becomes,
\begin{equation}
\frac{\partial}{\partial t}\mathcal{Y}(t,x)=i\left(\mathcal{H}+\frac{x_{1,t}'+g(t,z)}{\sqrt{2}}\mathcal{S}^c+\frac{x_{2,t}'}{\sqrt{2}}\mathcal{S}^a\right)\mathcal{Y}(t,x).\label{ssle3.5}
\end{equation}
By defining two colored noise terms $\xi_t=x_{1,t}'+g(t,z)$ and $\eta_t=ix_{2,t}$, we obtain the following stochastic equation:
\begin{eqnarray}
\frac{\partial}{\partial t}\mathcal{Y}(t,\xi,\eta)&=&\left(i\mathcal{H}+i\frac{\xi_t}{\sqrt{2}}\mathcal{S}^c+\frac{\eta_t}{\sqrt{2}}\mathcal{S}^a\right)\mathcal{Y}(t,\xi,\eta),\nonumber\\
\mathcal{M}_z\left\{\eta_t\eta_{s}\right\}&=&0,\nonumber\\
\mathcal{M}_z\left\{\xi_{t}\eta_s\right\}&=&2\theta(t-s)\,{\rm Im}[\alpha_T(t-s)],\nonumber\\
\mathcal{M}_z\left\{\xi_t\xi_{s}\right\}&=&2{\rm Re}[\alpha_T(|t-s|)].\nonumber\\\label{ssle4}
\end{eqnarray}
Here, $\theta(t-s)$ is the step function, which is 1 for $t>s$ and 0 for $t<s$; the value for $t=s$ is not important as Im$(\alpha_{T}(0))=0$. Each line of Eq.~(\ref{ssle4}) is a stochastic {\it c}-number Langevin equation of a stochastic basis operator $Y_m(t;\xi,\eta)$. Note that the noise terms in Eq.~(\ref{ssle4}) represent classical noise terms, therefore different from the quantum noise terms. By taking the noise average of these expectation values of the stochastic basis operators, we obtain the expectation values of the basis operators $\langle Y_m(t)\rangle=\mathcal{M}_z\{\langle Y_m(t;\xi,\eta)\rangle\}$. With these $\langle{Y_{m}(t)}\rangle$, the expectation value of any system operator can be conveniently calculated with the following relation:
\begin{equation}
\langle B(t)\rangle=\sum_lb_l\langle Y_m(t)\rangle.
\end{equation}

When the bath part of the operator $B(t)$ is not an identity operator, we just need to add an additional term in the noise average. Consequently, the expectation value can be calculated in a similar way,
\begin{eqnarray}
\langle B(t)\rangle&=&\mathcal{M}_z\left\{\sum_lb_l\langle{Y}_l(t;\xi,\eta)\rangle f(t,\zeta)\right\}.\nonumber\\\label{saoo2}
\end{eqnarray}

As we have mentioned, the effects of the bath are contained in the stochastic bath part of the operators instead of the artificial noise terms $z_{i,s_i}$, $i=1,2,3,$ and $4$. However, after the transformation in Eq.~(\ref{tron}), the information of the bath is absorbed into the noise terms $\xi_t$, $\eta_t$, and $f(t,\zeta)$. The influence of the bath on the system dynamics is described by $\xi_t$ and $\eta_t$; concerned bath operators are provided by $f(t,\zeta)$; other details of the bath are eliminated as in the master equation.
\section{The dynamics of two-level systems}
\label{example}
In this section, we will calculate the dynamics of different two-level systems as applications of our method. Two-level systems are the simplest kind of multi-level systems, and have a wide range of applications (see, e.g.,~\cite{mtt,uosct5,uosct2}). Meanwhile, the Langevin equation of a two-level system can also have the problem of nonlinear time-non-local terms. We will consider the coupling energy $\langle H_{\rm I}(t)\rangle$ and the bath displacement $\langle x(t)\rangle\equiv\langle\sum_{k}(g_k^*a_k^{\dag}(t)+g_ka_k(t))\rangle$ as two simple cases of quantities which contain bath operators. Note that the bath displacement has units of frequency instead of length, because it is multiplied by the coupling coefficient. Such a displacement can better reveal the influence of the bath on the system. The stochastic expectation values of the system part can be obtained from Eq.~(\ref{ssle4}), so we will directly come to the bath part. According to Eq.~(\ref{hamiltonian}) and Eq.~(\ref{pebot2}), we have
\begin{eqnarray}
\langle H_{\rm Ib}(t;z)\rangle&=&\left\langle\sum_{k}\left[g_k^*a_k^{\dag}(t;z)+g_ka_k(t;z)\right]\right\rangle\nonumber\\
                               &=&\sum_{k}\left[g_k^*\mathcal{A}_k(t;z)+g_k\mathcal{A}_k^*(t;z)\right]\langle I_{\rm b}(t;z)\rangle.\nonumber\\\label{tlsbp}
\end{eqnarray}
Equation (\ref{tlsbp}) can be evaluated with Eq.~(\ref{gsooad}) and Eq.~(\ref{seobu}),
\begin{eqnarray}
\langle H_{\rm Ib}(t;z)\rangle&=&\zeta_t\langle I_{\rm b}(t;z)\rangle,\nonumber\\
\zeta_t&=&\sqrt{2}\int_0^tds\left[(z_{1,s}+iz_{2,s})\tilde{\alpha}^*(t-s)\right.\nonumber\\
         &&\left.+(z_{3,s}-iz_{4,s})\tilde{\alpha}(t-s)\right],\nonumber\\
\tilde{\alpha}(t-s)&=&\alpha_T(t-s)-\frac{1}{2}\alpha(t-s).\nonumber\\\label{tlsbpe}
\end{eqnarray}
From Eqs.~(\ref{tlsbpe}), we can find that the contribution of the bath part can be represented by a new noise term $\zeta_{t}$. The correlation among $\zeta_t$, $\xi_t$, and $\eta_t$ is described through
\begin{eqnarray}
\mathcal{M}_z\{\zeta_t\zeta_s\}&=&0,\nonumber\\
\mathcal{M}_z\{\zeta_t\xi_s\}&=&\theta(t-s)2\sqrt{2}\,{\rm Re}\left[\alpha_T(t,s)-\frac{1}{2}\alpha(t,s)\right],\nonumber\\
\mathcal{M}_z\{\zeta_t\eta_s\}&=&\theta(t-s)2\sqrt{2}\,{\rm Im}\left[\alpha_T(t,s)-\frac{1}{2}\alpha(t,s)\right].\nonumber\\
\end{eqnarray}
Then, the coupling energy can be calculated with Eq.~(\ref{saoo2}) by introducing an additional noise to the noise average,
\begin{eqnarray}
\langle H_{\rm I}(t)\rangle&=&\mathcal{M}_z\left\{\langle S_{\rm s}(t;\xi,\eta)\rangle \zeta\right\},\nonumber\\\label{EoC}
\end{eqnarray}
where, $S_{\rm s}(t;\xi,\eta)$ is the system part stochastic expectation value of the coupling operator, $S(t)$, in Eq.~(\ref{hamiltonian}). The bath part of the bath displacement is just the same as the one of the coupling energy. Therefore, we have the following relation:
\begin{eqnarray}
\langle x(t)\rangle&=&\mathcal{M}_z\left\{\langle I_{\rm sys}(t;\xi,\eta)\rangle \zeta\right\}.\nonumber\\\label{dis}
\end{eqnarray}
We will mainly consider two kinds of spectral densities for the baths. The first spectral density is the Ohmic form with a Debye regulation (see, e.g.,~\cite{ods}):
\begin{eqnarray}
\sum_kg_kg_k^*\delta(\omega_k-\omega)&=&\frac{\Gamma\omega_c^2\omega}{\pi(\omega_c^2+\omega^2)},\label{spectrum}
\end{eqnarray}
where, $\Gamma$ is the coupling strength, and $\omega_c$ is the cut-off frequency. We will calculate several well-known simple models with this spectral density. In these models, we will express all the parameters in normalized units, as the ratio to the frequency of the two-level system $\omega_0$.

The second kind of spectral density we consider is the super-Ohmic spectrum with an exponential cut-off:
\begin{eqnarray}
\sum_kg_kg_k^*\delta(\omega_k-\omega)&=&\alpha\omega^3\exp\!\left({-\left(\frac{\omega}{\omega_c}\right)^2}\right),\label{spectrumso}
\end{eqnarray}
where the coupling strength is described by $\alpha$. This spectral function is frequently used in solid-state quantum dot systems, e.g., when coupled to acoustic phonons. We will first reproduce some known results, with our alternative techniques, and then obtain some new results. In addition, the bath displacement, which cannot be obtained with former methods, will also be calculated. The parameters in this part will be expressed with units of picosecond, like other works modeling real quantum dot systems~\cite{papi1,papi2}. In our numerical calculations, the noise functions are generated with the FFTW3 pack~\cite{fft}.
\subsection{Pure dephasing model}
The first model we consider is the so-called pure dephasing model:
\begin{eqnarray}
H_{\rm sys}=\frac{\omega_0}{2}\sigma_z,~~~S=\sigma_z.\label{exampdm}
\end{eqnarray}
We use such an analytically solvable model to check our method. According to Eq.~(\ref{ssle4}), the stochastic {\it c}-number Langevin equations of the pure dephasing model are
\begin{eqnarray}
\frac{\partial}{\partial  t}\langle\sigma_{xs}(t;\xi,\eta)\rangle&=&-\left(\omega_0+\sqrt{2}\xi_t\right) \langle\sigma_{ys}(t;\xi,\eta)\rangle,\nonumber\\
\frac{\partial}{\partial t}\langle\sigma_{ys}(t;\xi,\eta)\rangle&=&+\left(\omega_0+\sqrt{2}\xi_t\right)\langle\sigma_{xs}(t;\xi,\eta)\rangle,\nonumber\\
\frac{\partial}{\partial t}\langle\sigma_{zs}(t;\xi,\eta)\rangle&=&\sqrt{2}\eta_t\langle I_{\rm sys}(t;\xi,\eta)\rangle,\nonumber\\
\frac{\partial}{\partial t}\langle I_{\rm sys}(t;\xi,\eta)\rangle&=&\sqrt{2}\eta_t\langle\sigma_{zs}(t;\xi,\eta)\rangle.\nonumber\\\label{sclepd}
\end{eqnarray}
The noise averages of $\langle\sigma_{xs}(t;\xi,\eta)\rangle$, $\langle\sigma_{ys}(t;\xi,\eta)\rangle$, $\langle\sigma_{zs}(t;\xi,\eta)\rangle$, and $\langle I_{\rm sys}(t;\xi,\eta)\rangle$ correspond to the expectation values of three Pauli matrices and the identity operator. We assume the initial state of the system to be the eigenstate of $\sigma_x$ $|\psi_0\rangle=\frac{1}{\sqrt{2}}(|e\rangle+|g\rangle)$, where $|g\rangle$ ($|e\rangle$) is the ground state (excited state). The initial state of the bath is assumed to be a thermal state. The coupling energy between the system and the bath is
\begin{eqnarray}
\langle H_{\rm I}(t)\rangle&=&\mathcal{M}_z\left\{\langle \sigma_{zs}(t;\xi,\eta)\rangle \zeta\right\}.
\end{eqnarray}
From the analytical solution of this model, we obtain the expectation values of the following quantities,
\begin{eqnarray}
\langle \sigma_x(t)\rangle&=&\cos(\omega_0t)e^{-4\int_0^tds\int_0^sds_1{\rm Re}[\alpha_T(s-s_1)]}\langle \sigma_x(0)\rangle,\nonumber\\
\langle \sigma_y(t)\rangle&=&\sin(\omega_0t)e^{-4\int_0^tds\int_0^sds_1{\rm Re}[\alpha_T(s-s_1)]}\langle \sigma_x(0)\rangle,\nonumber\\
\langle H_{\rm I}(t)\rangle&=&2\int_0^t\!\!ds\,{\rm Im}[\alpha(t-s)].\label{anar}
\end{eqnarray}
With Eq.~(\ref{anar}), the stochastic results can be compared with the analytical results. We first show the results at low temperature $\beta=1000$ ($\beta$ has units of $1/\omega_0$).
\begin{figure*}
\includegraphics[width=6.5in]{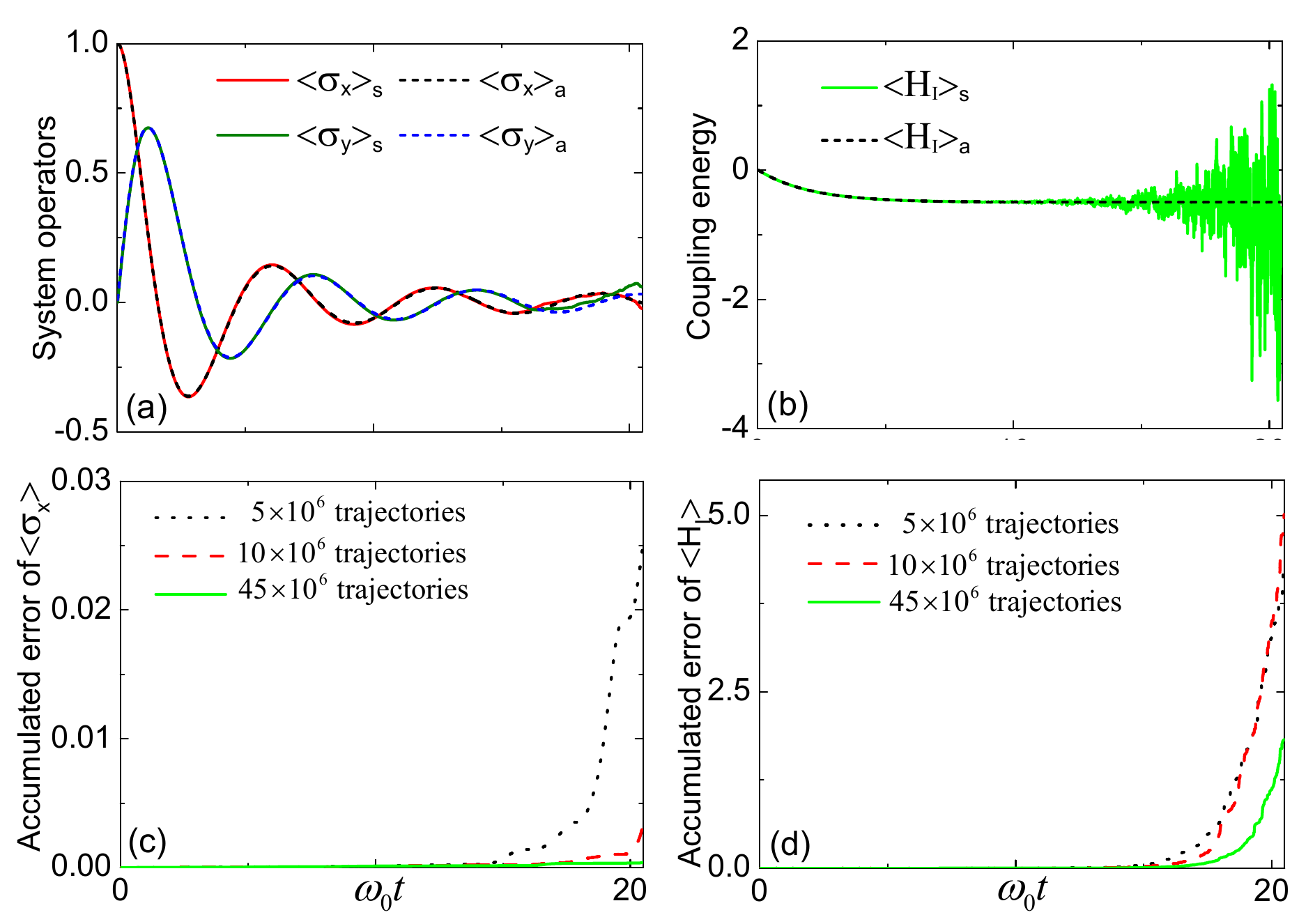}
\caption{Results for the pure dephasing model. Here, the cut-off frequency is $\omega_c=0.5$, and the coupling strength $\Gamma=1$. In (a) and (b), the stochastic results are calculated taking an average over $45\times10^{6}$ trajectories, and the results of the stochastic method and the analytical expression are denoted by the subscript $s$ and $a$, respectively. The accumulated errors are shown in (c) and (d).}
\label{fig1}
\end{figure*}

From Fig.~\ref{fig1}, we can see that the converged part of the stochastic results (solid curves) agree well with the analytical results (dotted curves). Figure~\ref{fig1}(a) shows the expectation values of the system operators; there, the stochastic results are nearly converged except for some tiny deviations at long times. However, the result in Fig.~\ref{fig1}(b), which shows the coupling energy, is not that good. For comparatively long times, $\omega_0t>15$, the stochastic result oscillates severely around the analytical result. Also note that the interaction energy agrees well with the analytical even for short times, so the approximation in getting the bath operators is reliable. Then, we introduce the accumulated errors of $\langle\sigma_x\rangle$ and $\langle H_{\rm I}\rangle$ to characterize the convergence of our method. The accumulated error of some operator $B(t)$ is calculated from
\begin{eqnarray}
{\rm Accumulated \ error} = \int_0^t\!\!ds\left[\langle B(s)\rangle_{\rm a}-\langle B(s)\rangle_{\rm s}\right]^2.\nonumber\\
\end{eqnarray}
The $\langle B(s)\rangle_{\rm a}$ and $\langle B(s)\rangle_{\rm s}$ correspond to the analytical result and the stochastic result, respectively. As shown in Fig.~\ref{fig1}(c), the error decreases significantly with the number of trajectories for the system operator $\sigma_x$. However, the calculation of the coupling energy is much more difficult. According to the results in Fig.~\ref{fig1}(c), doubling the number of trajectories (black dotted curve and red dashed curve) makes only a few percent difference in the error. Only by increasing the number of the trajectories from $5\times10^{6}$ (black dotted curve) to $45\times10^{6}$ (green solid curve), we can reduce the error slightly. Therefore, the converging properties of the system operators and the bath ones are quite different.
\begin{figure}
\includegraphics[width=3.3in]{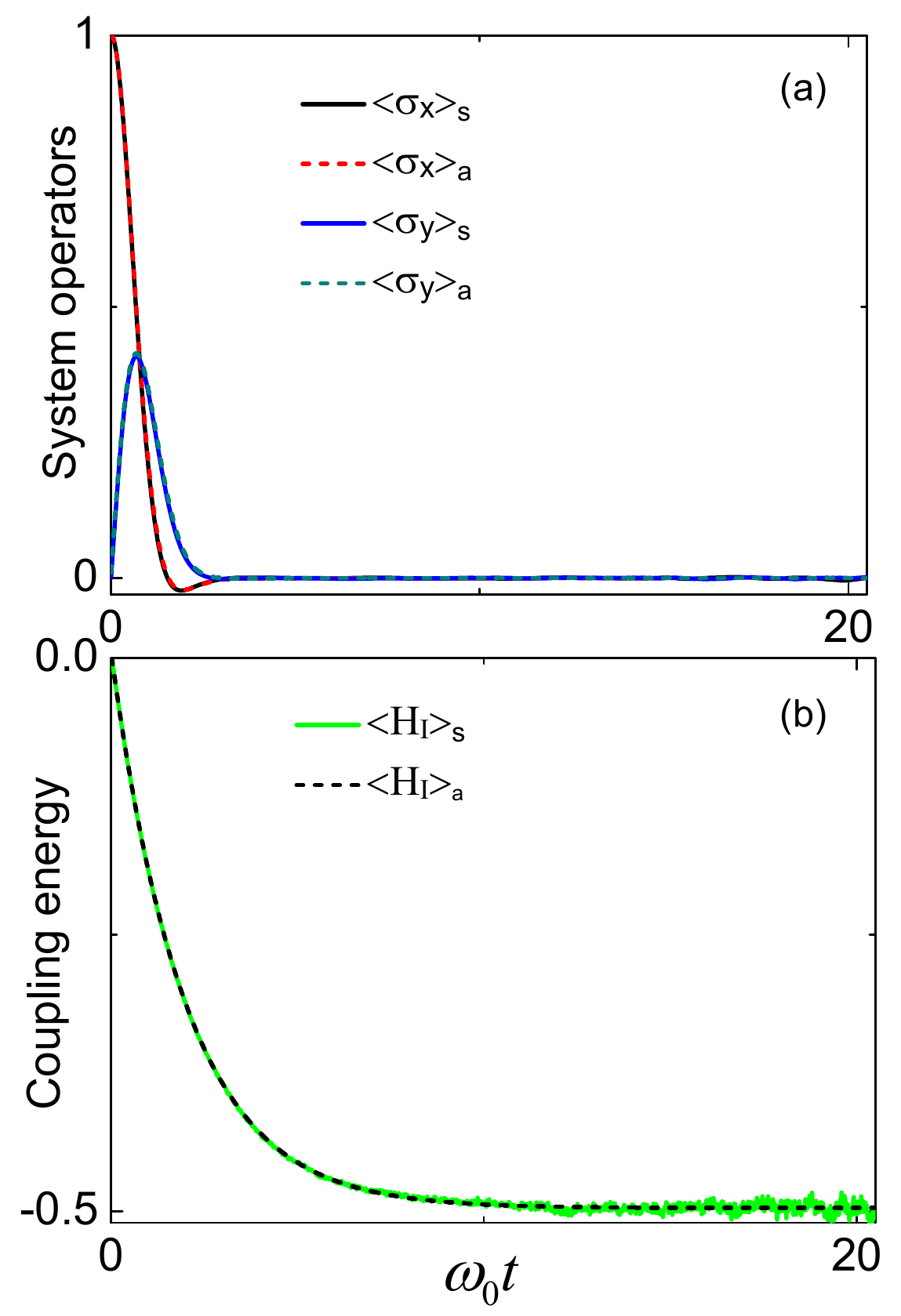}
\caption{Expectation values of different operators. The cut-off frequency is $\omega_c=0.5$, and the coupling strength $\Gamma=1$. The stochastic results are calculated taking an average over $5\times10^{6}$ trajectories. (a) Comparison between the stochastic and analytical results for the system operators. (b) Comparison between the stochastic and analytical results for coupling energy. }
\label{fig2}
\end{figure}

Next, we come to the case of high temperature $\beta=1$. Compared with Fig.~\ref{fig1}, the convergence of the results in Fig.~\ref{fig2} is much better. This is not surprising because the system is closer to the classical limit at high temperature. However, the convergence of the coupling energy $\langle H_{\rm}(t)\rangle$, shown in Fig.~\ref{fig2}(b), is still not as good as the system operators, shown in Fig.~\ref{fig2}(a).

\subsection{Pure dephasing model with classical field control}
We now consider a situation with classical field control. In addition to the original Hamiltonian in Eq.~(\ref{exampdm}), we will add a classical control field Hamiltonian
\begin{eqnarray}
H_{\rm c}=C(t)\sigma_y.
\end{eqnarray}
The coefficient $C(t)$ describes the shape of the control field, and we choose the ideal $\pi$-pulses as the control field. The separation between two pulses is $2/\omega_0$.
Thus the coefficient $C(t)$ can be written as
\begin{eqnarray}
C(t)=\sum_n\frac{\pi}{2}\;\delta\!\left(t-\frac{2n}{\omega_0}\right).
\end{eqnarray}
The derivation in Sec.~\ref{secsle} is based on time-independent Hamiltonians, but this formalism also works for time-dependent Hamiltonians (see Appendix D). The numerical results are shown in Fig.~\ref{fig5}.
\begin{figure}
\includegraphics[width=3.3in]{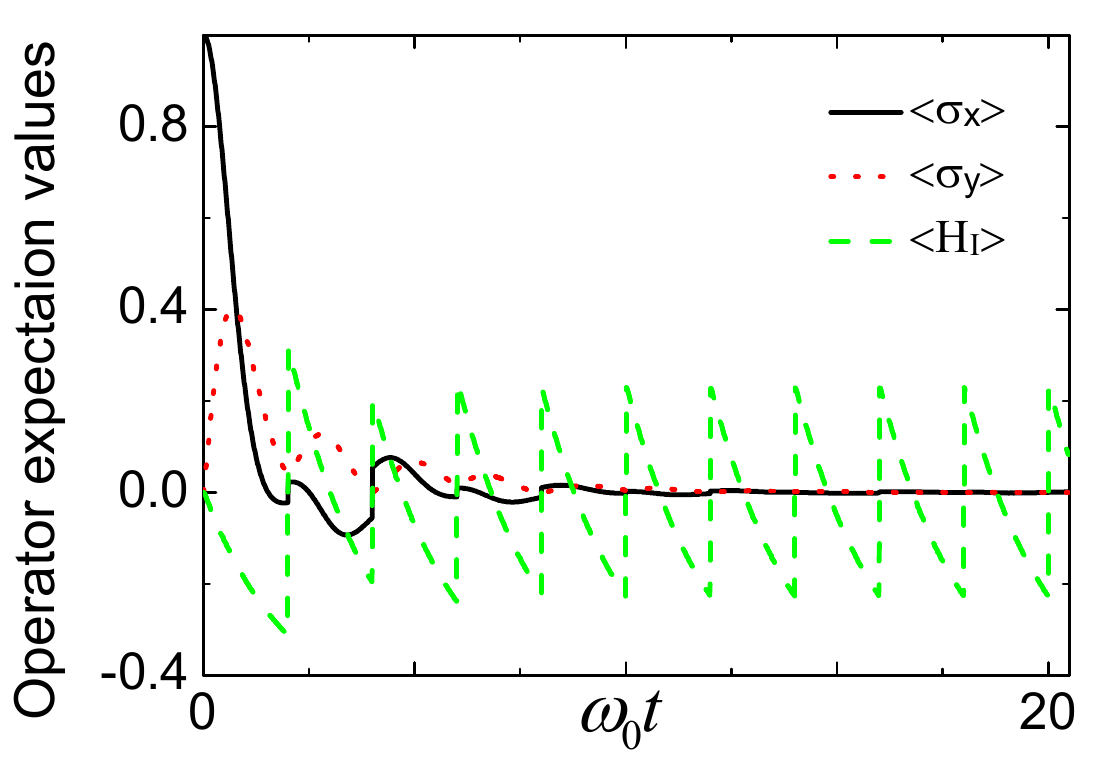}
\caption{Expectation values of different operators with classical control. The cut-off frequency is $\omega_c=0.5$, the coupling strength is $\Gamma=1$, and the inverse of the temperature is $\beta=1$. The stochastic results are calculated with the average over $40\times10^{6}$ trajectories. The control field is ideal $\pi$-pulses in y-direction. }
\label{fig5}
\end{figure}

In Fig.~\ref{fig5}, our results correctly show the effects of the control field. Compared with the results without control field in Fig.~\ref{fig2}, the coherence time is much longer. At long times, the expectation values of the system operators become the steady values, but the coupling energy cannot become a steady value under the influence of the control pulses. The sudden changes in the Fig.~\ref{fig5} correspond to the control pulses. The $\langle\sigma_z(t)\rangle$ is not shown as it is a conserved quantity for our initial state.
\subsection{Spin-Boson model}
The second model we consider is the spin-boson model, which, unlike the pure dephasing mode, cannot be solved analytically. The spin-boson model is very important as it can describe light-matter interaction problems and double-well potential problems. The system Hamiltonian and coupling operator of the spin boson model are,
\begin{eqnarray}
H_{\rm sys}=\frac{\omega_0}{2}\sigma_z,~~~S=\sigma_x.
\end{eqnarray}
The stochastic {\it c}-number Langevin equations of the system operators are as follows,

\begin{eqnarray}
\frac{\partial}{\partial  t}\langle\sigma_{xs}(t;\xi,\eta)\rangle&=&-\omega_0\langle\sigma_{ys}(t;\xi,\eta)\rangle+\sqrt{2}\eta_t\langle I_s(t;\xi,\eta)\rangle,\nonumber\\
\frac{\partial}{\partial t}\langle\sigma_{ys}(t;\xi,\eta)\rangle&=&\omega_0\langle\sigma_{xs}(t;\xi,\eta)\rangle-\sqrt{2}\xi_t\langle\sigma_{zs}(t;\xi,\eta)\rangle,\nonumber\\
\frac{\partial}{\partial t}\langle\sigma_{zs}(t;\xi,\eta)\rangle&=&\sqrt{2}\xi_t\langle\sigma_{ys}(t;\xi,\eta)\rangle,\nonumber\\
\frac{\partial}{\partial t}\langle I_s(t;\xi,\eta)\rangle&=&\sqrt{2}\eta_t\langle\sigma_{xs}(t;\xi,\eta)\rangle.\label{sclesb}
\end{eqnarray}
The coupling energy can be estimated with
\begin{eqnarray}
\langle H_{\rm I}(t)\rangle&=&\mathcal{M}_z\left\{\langle \sigma_{xs}(t;\xi,\eta)\rangle \zeta\right\},
\end{eqnarray}
and we assume the bath to be in the thermal state, while the system in the excited state $|e\rangle$.
\begin{figure}
\includegraphics[width=3.5in]{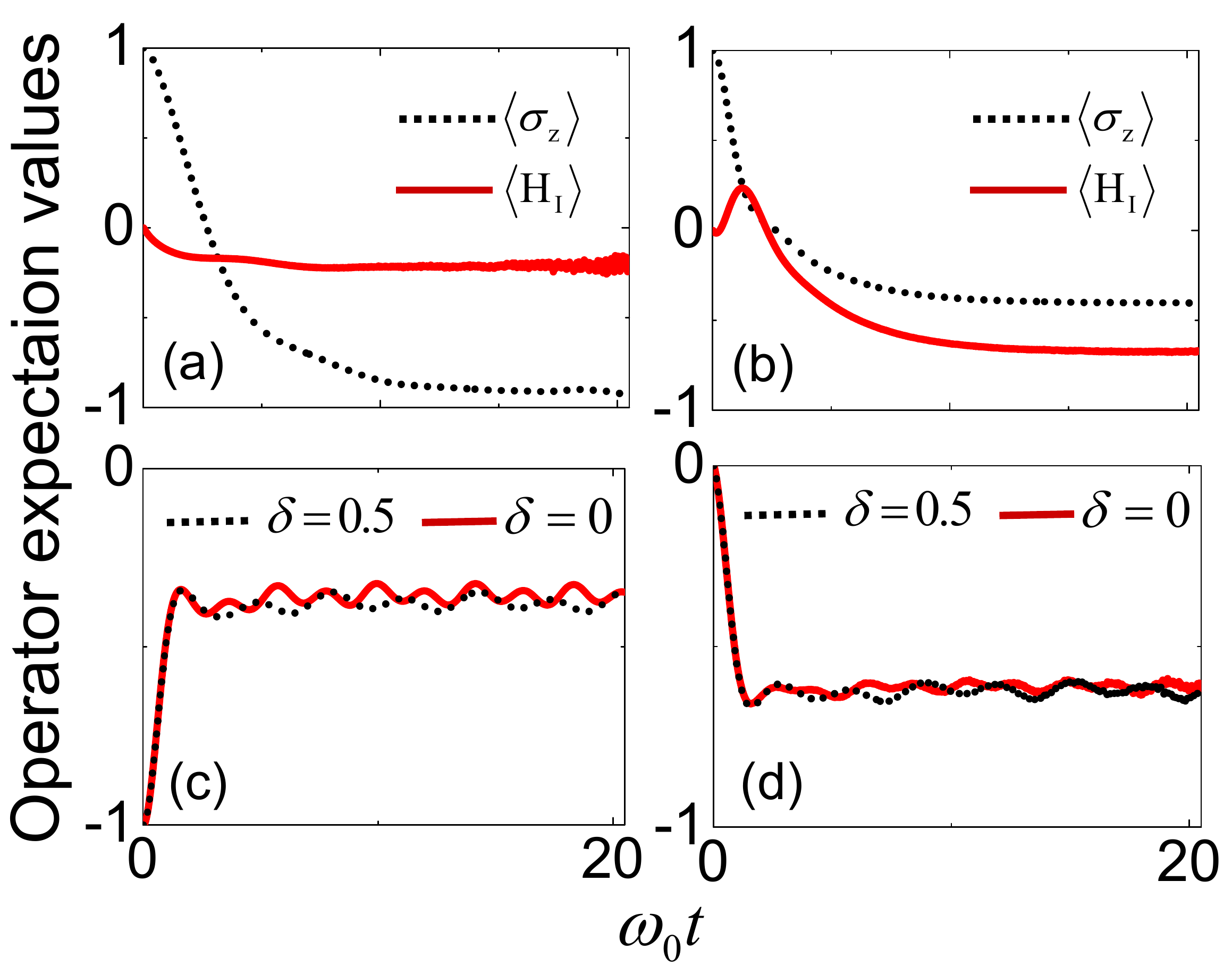}
\caption{Expectation values of different operators. The cut-off frequency is $\omega_c = 0.5$, and the coupling strength is $\Gamma=1$. The stochastic results are calculated with the average over $40\times10^{6}$ trajectories for (a) and (b). (a) The results at the low temperature $\beta= 1000$. (b) The results at hight temperature  $\beta=1$. The pumped cases, shown in (c)
and (d), are calculated at high temperature  $\beta=1$ with $10\times10^{6}$ trajectories. The expectation values of $\sigma_z$ are shown in (c), and the coupling energies are shown in (d). }
\label{fig3}
\end{figure}

The convergence of the results, as shown in Fig.~\ref{fig3}, is comparatively good, and there is only some small fluctuations at the end of the red curve in Fig.~\ref{fig3}(a). Similar to the pure dephasing model, the coupling energy at low temperature is hard to calculate. The coupling energy decreases monotonically when the temperature is low (Fig.~\ref{fig3}(a)). In the high temperature case (Fig.~\ref{fig3}(b)), the coupling energy increases first, then decreases to a steady value. As mentioned above, our calculation of the coupling energy can be unreliable for short times, so this difference should be treated carefully. The steady-state value of the interaction energy at higher temperatures is much lower than the low-temperature one. Then we add an optical pump term to the system,
\begin{eqnarray}
H_{\rm p}&=&\frac{\Omega}{2}\sin((\omega_0+\delta)t)\sigma_x,
\end{eqnarray}
The peak Rabi frequency is set to $\Omega=0.5$ (in units of $\omega_0$), and we consider different detunings. We assume the initial state of the system its ground state. From Figs. \ref{fig3}(c) and \ref{fig3}(d), we can find that the influence of the detuning is insignificant. This means that the effects of the pump are suppressed by the bath. However, note that the pump here only differs from $S=\sigma_x$, which describes the coupling to the bath, by a coefficient. We will show in the next example that the situation can be different when $S=\sigma_z$.

\subsection{Optically pumped quantum dot system}
Next we consider an optically pumped quantum dot system, which has many applications in solid state quantum optics. The system considered here is a quantum dot with a pump and electron-phonon coupling, which has been studied in various experiments (e.g.,~\cite{qdots1,qdots2}). The Hamiltonian under the rotating frame is
\begin{eqnarray}
H_{\rm sys}&=&\frac{\delta}{2}\sigma_z+\frac{\Omega(t)}{2}\sigma_x,~~~~S=\frac{\sigma_z}{2}.\nonumber\\
\end{eqnarray}
Here, the detuning of the pump is $\delta$, and the Rabi frequency of the pump is $\Omega(t)$. The spectrum of the bath is of super-Ohmic form. We first consider the case of on-resonant pumping $\delta=0$. The parameters used here are from reference~\cite{polatran3}. In addition to the population of the excited state, which is ${(1+\langle\sigma_z\rangle)}/{2}$, we also calculate the bath displacement induced by the system $\langle x(t)\rangle=\sum_k\langle(g_k^*a_k^{\dag}(t)+g_ka_k(t))\rangle$.
\begin{figure}
\includegraphics[width=3.5in]{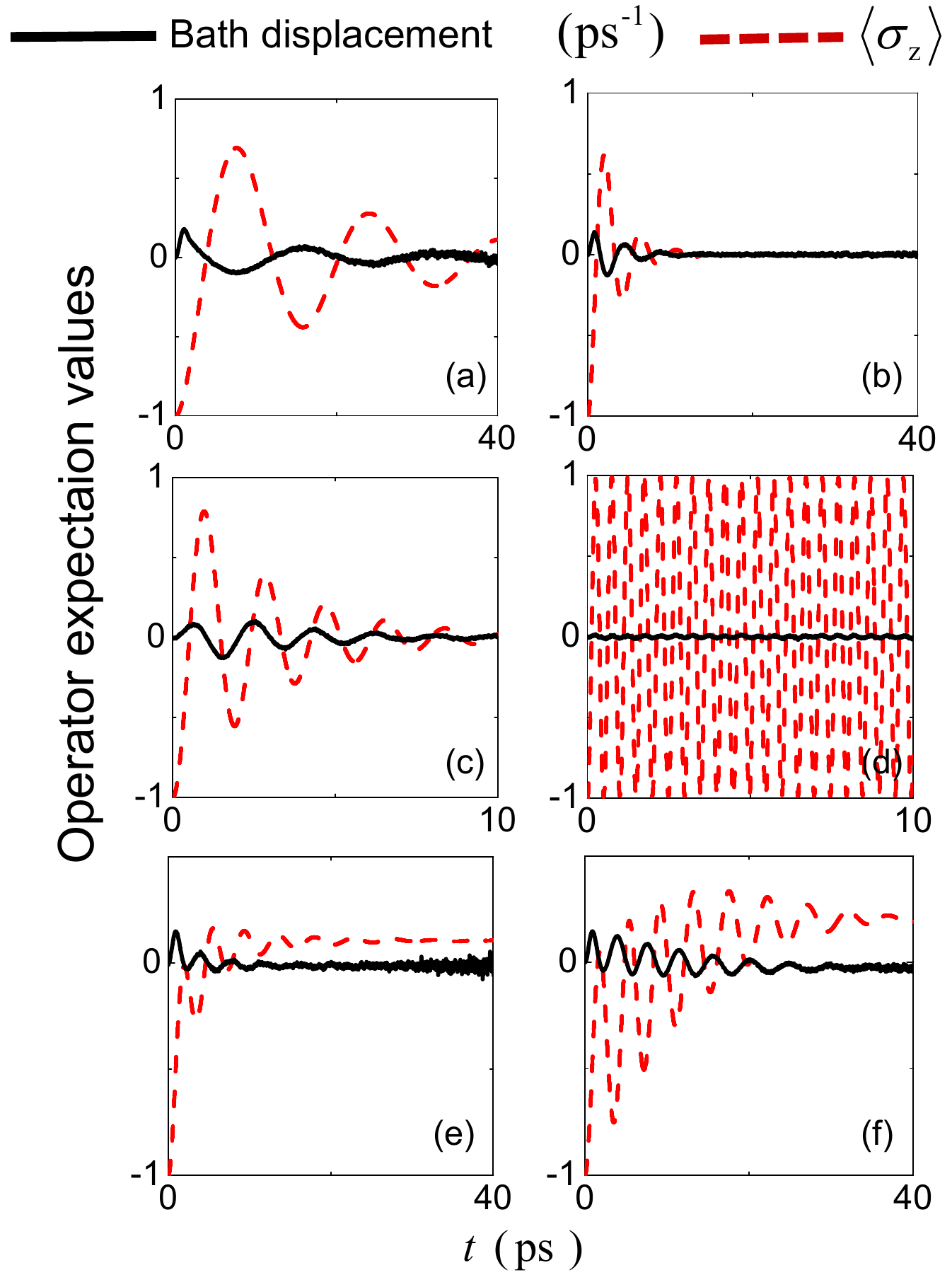}
\caption{Quantum dot systems with different pump intensity. The displacement of the bath is calculated using $\langle x(t)\rangle=\sum_k\langle(g_k^*a_k^{\dag}(t)+g_ka_k(t))\rangle$. The cut-off frequency is $\omega_c=2.2~{\rm ps}^{-1}$, and the coupling strength is $\alpha=0.027~{\rm ps}^{-2}$. The temperatures are $T=50~{\rm K}$ for (a)-(e), and the temperature for (f) is $4.2~{\rm K}$. The detunings for (a)-(d) are $\delta=0$, and for (e) and (f), the detunings are $\delta=-1.26~{\rm ps}^{-1}$. The stochastic results are calculated with the average over $1\times10^6$ trajectories. The pump intensities are $\Omega=\pi/6~{\rm ps}^{-1}$ for (a), $\Omega=\pi/2~{\rm ps}^{-1}$ for (b), $\Omega=\pi~{\rm ps}^{-1}$ for (c), and $\Omega=4\pi~{\rm ps}^{-1}$ for (d). For (e) and (f), the pump intensities are time dependent, $\Omega(t)=1.28\exp(-(t/\tau)^2)~{\rm ps}^{-1}$ with $\tau=20.2~{\rm ps}$, corresponding to the pulse area $\Theta=14.6\pi$.}
\label{fig6}
\end{figure}
From the red dashed curves in Figs.~\ref{fig6}(a)-(c), our calculations agree well with other results~\cite{polatran3}; namely, the population oscillations for different pump Rabi frequencies, and there is bath-induced damping. The bath displacements (black solid curves) can provide more insights into the bath dynamics and coupling. When the pump intensities are moderate [Figs.~\ref{fig6}(b) and~\ref{fig6}(c)], the bath displacements are comparatively large and oscillate with the same frequencies of the oscillations of $\langle\sigma_z\rangle$. The bath displacement is still large if the pump is weak [Fig.~\ref{fig6}(a)], but the oscillation is not resonant with the system. For very strong pumping [Fig.~\ref{fig6}(d)], the bath displacement is very small, and there is little bath-induced damping; this is consistent with other findings~\cite{polatran3}. The bath displacement here agrees with the estimation of the variational polaron transform master equation, namely, nearly zero bath response at large pump Rabi frequency. In addition, our
method can also provide a clear picture of the bath response when the Rabi frequency of the pump is close to the cut-off frequency. In such a case, the bath displacement and the system population always oscillate at the same frequency, and the dissipation reaches its maximum.

Next we consider the time-dependent pump intensity \cite{polatran8}. The parameters of the bath are the same as the ones in the time-independent case. We set the detuning to be $\delta= -1.26~{\rm ps}^{-1}$, and the pump intensity to be $\Omega(t) = \Omega_0 \exp(-(t/\tau )^2)$ with $\Omega_0=1.28~{\rm ps}^{-1}$ and $\tau=20.2~{\rm ps}^{-1}$ corresponding to the pulse area $\Theta\equiv\sqrt{\pi}\Omega_0\tau=14.6\pi$. As shown in Figs.~\ref{fig6}(e) and~\ref{fig6}(f), the populations (red dashed curves) are inverted with negative detunings. Also, the steady values of the bath displacements are not zero in these cases. This is consistent with other works and experiments~\cite{papi1,papi2,papi3}, which shows that the phonon dissipation can assist to invert the population of a pumped quantum dot when the detuning is negative.

In this section, we have calculated the expectation values of different operators in two-level systems. Although our results do not totally converge in Fig.~\ref{fig1}(b), the coupling energy has reached its steady-state value before the results diverge. For other examples, the trajectory numbers are sufficient to obtain convergence. The numerical efficiency of our approach can change with the problem studied. Here, we summarize the numbers of trajectories used for different results in the following table:
\begin{table}[htbp]
\caption{Numbers of trajectories for different results}
\begin{tabular}{|c|c|}
\hline
Pure dephasing model (low temperature)&$4.5\times10^7$\\
\hline
Pure dephasing model (high temperature)&$5\times10^6$\\
\hline
Pure dephasing model with control&$4\times10^7$\\
\hline
Spin-boson model&$4\times10^7$\\
\hline
Spin-boson model with pump&$10^7$\\
\hline
Quantum dot with optical pump&$10^6$\\
\hline
\end{tabular}
\end{table}

We have shown that our method can obtain the expectation values of operators which contain bath operators. Also, our method can deal with both low and high temperature cases. However, the low temperature or the non-trivial bath parts can increase the calculation costs.

\section{conclusions}
\label{conclusion}
We have developed a stochastic {\it c}-number Langevin equation method to access both the system information and the bath information. The problem of the nonlinear time-non-local terms in Langevin equations is avoided by formally dividing all the operators into system parts and bath parts with auxiliary stochastic fields (noise). As a Heisenberg-Langevin method, our approach can conveniently access the bath information. Such information about the bath can be quite different even when the dynamics of Pauli matrices [$\sigma_x(t)$, $\sigma_y(t)$, and $\sigma_z(t)$] are similar. In addition, this method is not limited to certain bath spectra or temperature in spite of the increased computing cost at low temperatures. We have also applied our method to several cases. For example, our equations work well in different well known simple models, like the pure dephasing model and the spin-boson model, and can compute both system and bath quantities. We have also reproduced some existing results in a pumped quantum dot system, and extended our model into a range of validity where typically these methods fail, In addition, the bath displacement, which is difficult to obtain with former approaches, can be calculated with our methodology. Finally, we stress that our method can be applied to problems like closed many-body systems or quantum thermodynamics.\\\\
%\section*{ACKNOWLEDGMENTS}

\begin{acknowledgments}
We are grateful to Prof. Wei-Min Zhang and Chris Gustin for their instructive comments. Z.Y.Z. is supported by the Japan Society for the Promotion of Science (JSPS) Foreign Postdoctoral Fellowship No. P17821. Y.A.Y. is supported by the National Natural Science
Foundation of China (Grant No. 21373064).  J.Q.Y. is partially supported by the National Key Research and Development Program of China (Grant No. 2016YFA0301200) and the National Natural Science Foundation of China (Grant No. 11774022 and No. U1801661).  F.N. is
supported in part by the: MURI Center for Dynamic Magneto-Optics via the Air Force Office of Scientific Research (AFOSR) (FA9550-14-1-0040), Army Research Office (ARO) (Grant No. W911NF-18-1-0358), Asian Office of Aerospace Research and Development (AOARD)
(Grant No. FA2386-18-1-4045), Japan Science and Technology Agency (JST) (via the Q-LEAP program, and the CREST Grant No. JPMJCR1676), Japan Society for the Promotion of Science (JSPS) (JSPS-RFBR Grant No. 17-52-50023, and JSPS-FWO Grant No. VS. 059.
18N), the RIKEN-AIST Challenge Research Fund, and the John Templeton Foundation. S.H. thanks the Natural Sciences and Engineering Research Council of Canada for funding.
\end{acknowledgments}
%\vspace*{-0.3in}
\begin{widetext}
\appendix
\section{Stochastic bath evolution operator}
\label{AppA}
The solution of Eq.~(\ref{sbo2}) can be obtained with the Magnus expansion,
\begin{eqnarray}
{U}_{Ib}(t;z)&=&\exp\!\left(-i\frac{1}{\sqrt{2}}\sum_k\int_0^t\!\! ds(iz_{1,s}+z_{2,s})(g_{k}e^{-i\omega_ks}a_{kb}+g_{k}^*e^{i\omega_ks}a^\dag_{kb})\right)\times\nonumber\\
                    &&\exp\!\left(-\sum_k\int_0^t\!\!ds_1\int_0^{s_1}\!\!ds_2\frac{1}{4}(iz_{1,s_1}+z_{2,s_1})(iz_{1,s_2}+z_{2,s_2})(g_kg_k^*e^{-i\omega_k(s_1-s_2)}-g_kg_k^*e^{i\omega_k(s_1-s_2)})\right).
\label{mesbe}
\end{eqnarray}
Then we will separate the annihilation operators $a_k$ and creation operators $a_k^{\dag}$ in Eq.~(\ref{mesbe}) with the Baker-Campbell-Hausdorff formula,
\begin{eqnarray}
{U}_{Ib}(t;z)&=&\exp\!\left(-i\frac{1}{\sqrt{2}}\int_0^t\!\! ds(iz_{1,s}+z_{2,s})\sum_kg_{k}^*e^{i\omega_ks}a^\dag_{kb}\right)\exp\!\left(-i\frac{1}{\sqrt{2}}\int_0^t\!\! ds(iz_{1,s}+z_{2,s})\sum_kg_{k}e^{-i\omega_ks}a_{kb}\right)\times\nonumber\\
              &&\exp\!\left(-\frac{1}{2}\int_0^t\!\!ds_1ds_2(iz_{1,s_1}+z_{2,s_1})(iz_{1,s_2}+z_{2,s_2})\alpha(s_1-s_2)\right).\nonumber\\
\end{eqnarray}
\section{The generator of the bath operators}
\label{AppB}
By taking the functional variation on the noise term in Eq.~(\ref{sbo2}), we can obtain,
\begin{eqnarray}
\frac{\delta}{\delta z_{4,s_1}}U_{Ib}(t;z_{3,\tau},z_{4,\tau})&=&-i\sum_k\frac{g_k^*e^{i\omega_ks_1}}{\sqrt{2}}a^{\dag}_{kb}U_{Ib}(t;z_{3,\tau},z_{4,\tau})-i\frac{g_ke^{-i\omega_ks_1}}{\sqrt{2}}U_{Ib}(t;z_{3,\tau},z_{4,\tau})a_{kb}\nonumber\\
                                          &&-\frac{1}{2}\int_0^{s_1}\!\!ds_2\alpha(s_1-s_2)(iz_{3,s_2}+z_{4,s_2})U_{Ib}(t;z_{3,\tau},z_{4,\tau})\nonumber\\
                                          &&-\frac{1}{2}\int_{s_1}^t\!\!ds_2\alpha(s_2-s_1)(iz_{3,s_2}+z_{4,s_2})U_{Ib}(t;z_{3,\tau},z_{4,\tau}).\nonumber\\
\end{eqnarray}
For sufficiently long times, we can obtain the creation operator of the $k$th bath mode with Fourier transforms. Note that this approximation may introduce some error in the short-time limit. We have
\begin{eqnarray}
a_{kb}^{\dag}U_{Ib}(t;z_{3,\tau},z_{4,\tau})&=&\frac{i\sqrt{2}}{g_k^*}\int_0^t\!\!ds_1e^{-i\omega_ks_1}\frac{\delta}{\delta z_{4,s_1}}U_{Ib}(t;z_{3,\tau},z_{4,\tau})\nonumber\\
&&+\frac{i}{g_k^*\sqrt{2}}\int_0^t\!\!ds_1\int_0^{s_1}\!\!ds_2(iz_{3,s_2}+z_{4,s_2})e^{-i\omega_ks_1}\alpha(s_1,s_2)U_{Ib}(t;z_{3,\tau},z_{4,\tau})\nonumber\\
&&+\frac{i}{g_k^*\sqrt{2}}\int_0^t\!\!ds_1\int_{s_1}^t\!\!ds_2(iz_{3,s_2}+z_{4,s_2})e^{-i\omega_ks_1}\alpha^*(s_1,s_2)U_{Ib}(t;z_{3,\tau},z_{4,\tau})\nonumber\\
                                            &\equiv&\mathcal{A}_k(t;z_{3,\tau},z_{4,\tau})U_{Ib}(t;z_{3,\tau},z_{4,\tau}).\nonumber\\\label{gc}
\end{eqnarray}
The annihilation operator $a_k$ can be easily obtained by taking the transpose conjugate of Eq.~(\ref{gc}).
\section{The stochastic expectation value of bath identity operator $\langle I_b(t;z)\rangle$}
\label{AppC}
Let us now calculate the average value of the stochastic bath identity operator:
\begin{eqnarray}
\langle I_b(t;z)\rangle&=&{\rm Tr}\left[\rho_b(0)U_{Ib}^{\dag}(t;z_{1,\tau},z_{2,\tau})e^{i\sum_{k'}\omega_{k'} a^{\dag}_{k'b}a_{k'b}t}I_{b}(0)e^{-i\sum_{k}\omega_k a^{\dag}_{kb}a_{kb}t}U_{Ib}(t;z_{3,\tau},z_{4,\tau})\right]\nonumber\\
                       &=&\exp\!\left(i\frac{1}{\sqrt{2}}\int_0^t \!\!ds(-iz_{1,s}+z_{2,s})\sum_kg_{k}^*e^{i\omega_{k}s}a_{kb}^{\dag}\right)\exp\!\left(i\frac{1}{\sqrt{2}}\int_0^t\!\! ds(-iz_{1,s}+z_{2,s})\sum_kg_{k}e^{-i\omega_{k}s}a_{kb}\right)\times\nonumber\\
&&\exp\!\left(-i\frac{1}{\sqrt{2}}\int_0^t\!\! ds(iz_{3,s}+z_{4,s})\sum_kg_{k}^*e^{i\omega_ks}a^\dag_{kb}\right)\exp\!\left(-i\frac{1}{\sqrt{2}}\int_0^t\!\! ds(iz_{3,s}+z_{4,s})\sum_kg_{k}e^{-i\omega_ks}a_{kb}\right)\times\nonumber\\
&&\exp\!\left(-\int_0^t\!\!ds_1\int_0^{s_1}\!\!ds_2\frac{1}{2}(-iz_{1,s_1}+z_{2,s_1})(-iz_{1,s_2}+z_{2,s_2})\alpha^*(s_1-s_2)\right)\times\nonumber\\
&&\exp\!\left(-\int_0^t\!\!ds_1\int_0^{s_1}\!\!ds_2\frac{1}{2}(iz_{3,s_1}+z_{4,s_1})(iz_{3,s_2}+z_{4,s_2})\alpha(s_1-s_2)\right).\nonumber\\\label{apcsuk}
\end{eqnarray}

The operators in Eq.~(\ref{apcsuk}) can be rearranged in normal order as follows,
\begin{eqnarray}
&&\exp\!\left(\sqrt{2}\int_0^t \!\!dsx^*_{1,s}\sum_kg_{k}^*e^{i\omega_{k}s}a_{kb}^{\dag}\right)\times\nonumber\exp\!\left(\sqrt{2}\int_0^t \!\!dsx^*_{1,s}\sum_kg_{k}e^{-i\omega_ks}a_{kb}\right)\times\nonumber\\
&&\exp\!{\left(\frac{1}{2}\int_0^t\!\! ds_1ds_2(-iz_{1,s_1}+z_{2,s_1})(iz_{3,s_2}+z_{4,s_2})\alpha(s_1-s_2)\right).}\nonumber\\
\end{eqnarray}
The $x^*_{1,s}$ is the complex conjugate of the noise term $x_{1,t}$ in Eq.~(\ref{ssle3}): Then, we can calculate the trace in Eq.~(\ref{apcsuk}).
\begin{eqnarray}
\langle I_{\rm b}(t;z)\rangle&=&\Pi_{k',k}\sum_{n_{k'},m_1,m_2}\langle n_{k'}|\frac{(1-e^{-\beta\omega_{k'}})e^{-\beta\omega_{k'}n_{k'}}}{m_1!m_2!}\left(\sqrt{2}\int_0^t\!\! dsx_{1,s}^*g_{k}^*e^{i\omega_{k}s}a_{kb}^{\dag}\right)^{m_1}\left(\sqrt{2}\int_0^t\!\! dsx_{1,s}^*g_{k}e^{-i\omega_ks}a_{kb}\right)^{m_2}|n_{k'}\rangle\times\nonumber\\
&&\exp\left(\int_0^t\!\!ds_1\int_0^{s_1}\!\!ds_2x_{1,s}^*\left[(z_{1,s_2}+iz_{2,s_2})\alpha^*(s_1,s_2)+(z_{3,s_2}-iz_{4,s_2})\alpha(s_1,s_2)\right]\right)\nonumber\\
&=&\Pi_k\sum_{n_k}(1-e^{-\beta\omega_k})e^{-\beta\omega_k{n_k}}\sum_{m\leq n_k}\frac{n_k!}{m!^2(n_k-m)!}\left(2\int_0^t\!\! ds_1ds_2x_{1,s_1}^*x_{1,s_2}^*g_{k}^*g_{k}e^{i\omega_{k}(s_1-s_2)}\right)^{m}\times\nonumber\\
&&\exp\left(\int_0^t\!\!ds_1\int_0^{s_1}\!\!ds_2x_{1,s}^*\left[(z_{1,s_2}+iz_{2,s_2})\alpha^*(s_1,s_2)+(z_{3,s_2}-iz_{4,s_2})\alpha(s_1,s_2)\right]\right)\nonumber\\
&=&\Pi_k\sum_{m}\frac{1}{m!}\left(2\int_0^t\!\! ds_1ds_2x_{1,s_1}^*x_{1,s_2}^*g_{k}^*g_{k}e^{i\omega_{k}(s_1-s_2)}\right)^{m}e^{-\beta\omega_km}\sum_{(n-m)\geq0}(1-e^{-\beta\omega_k})e^{-\beta\omega_k(n-m)}\frac{n!}{m!(n-m)!}\times\nonumber\\
&&\exp\left(\int_0^t\!\!ds_1\int_0^{s_1}\!\!ds_2x_{1,s}^*\left[(z_{1,s_2}+iz_{2,s_2})\alpha^*(s_1,s_2)+(z_{3,s_2}-iz_{4,s_2})\alpha(s_1,s_2)\right]\right)\nonumber\\
&=&\exp\left(\int_0^t\!\!ds_1\int_0^{s_1}\!\!ds_2x_{1,s_1}^*(z_{3,s_2}-iz_{4,s_2})\sum_kg_kg_k^*\left[\frac{1+e^{-\beta\omega_k}}{1-e^{-\beta\omega_k}}\cos(\omega_k(s_1-s_2))-i\sin(\omega_k(s_1-s_2))\right]\right)\times\nonumber\\
&&\exp{\left(\int_0^t\!\! ds_1\int_0^{s_1}\!\!ds_2x_{1,s_1}^*(z_{1,s_2}+z_{2,s_2})\sum_kg_kg_k^*\left[\frac{1+e^{-\beta\omega_k}}{1-e^{-\beta\omega_k}}\cos(\omega_k(s_1-s_2))+i\sin(\omega_k(s_1-s_2))\right]\right)}.\nonumber\\\label{apsevbu}
\end{eqnarray}
Now, we define the correlation function at temperature $T$ as
$$\alpha_T(t,s)=\sum_kg_kg_k^*\left[\frac{1+e^{-\beta\omega_k}}{1-e^{-\beta\omega_k}}\cos(\omega_k(s_1-s_2))-i\sin(\omega_k(s_1-s_2))\right].$$
Subsequently, the stochastic expectation value of the bath identity operator in Eq.~(\ref{apsevbu}) can be expressed in a compact form,
\begin{equation}
\langle I_b(t;z)\rangle=\exp\!\left(\int_0^t\!\!ds_1\int_0^{s_1}\!\!ds_2x_{1,s_1}^*\left[(z_{1,s_2}+iz_{2,s_2})\alpha_T^*(s_1,s_2)+(z_{3,s_2}-iz_{4,s_2})\alpha_T(s_1,s_2)\right]\right).
\end{equation}
\section{System with classical field control}
\label{AppD}
The derivations in Sec.~\ref{secsle} do not allow for a time-dependent Hamiltonian, because we have used the property that the Hamiltonian is unchanged during the time evolution. Now we provide the derivation of the time-dependent case. Consider a system with control field,
\begin{eqnarray}
H_{\rm sysc}(t)=H_{\rm sys}+\sum_{l}C_l(t)S_{l,\rm c}.\label{apch}
\end{eqnarray}
Here, the $H_{\rm sys}$ is the system Hamiltonian without control, $S_{l,\rm c}$ is a system operator, and $C_l(t)$ is the control function. Note that the Hamiltonian in Eq.~(\ref{apch}) is an approximate one. The original Hamiltonian should be
\begin{eqnarray}
H_{\rm origin}=H_{\rm sys}+\sum_{l}C_{l,\rm ext}S_{l,\rm c}+H_{\rm ext}.\label{apch2}
\end{eqnarray}
The control field is also governed by the quantum dynamics. The contribution of the external Hamiltonian $H_{\rm ext}$ is usually assumed to be very large (classical limit). Therefore, the evolution of the operator of the external control $C_{l,\rm ext}$ is only decided approximately by $H_{\rm ext}$. If we further omit the entanglement between the system and the external field, the operator of the external field $C_{l,\rm ext}$ can be substituted with its classical expectation value, so that
\begin{eqnarray}
H_{\rm sysc}(t)&=&H_{\rm sys}+{\rm Tr}_{\rm ext}\left\{\left(\sum_{l}C_{l,\rm ext}S_{l,\rm c}+H_{\rm ext}\right)\rho_{\rm ext}(t)\right\}\nonumber\\
               &=&H_{\rm sys}+\sum_{l}C_l(t)S_{l,\rm c}.\label{apch3}
\end{eqnarray}
The effective Hamiltonian in Eq.~(\ref{apch3}) is obtained in the Schr$\rm\ddot{o}$dinger picture, so it can
not be used directly in the Heisenberg picture. Now, we start from the original Hamiltonian in Eq.~(\ref{apch2}), which can be solved with our method:
\begin{eqnarray}
H_{\rm tot}&=&H_{\rm original}+\sum_{k}\omega_ka^{\dag}_ka_k+S\sum_{k}(g_k^*a_k^{\dag}+g_ka_k).\nonumber\\
\label{apcht1}
\end{eqnarray}
According to Eq.~(\ref{ssle1}), the stochastic equations of the external field operator $O_{\rm ext}$ and system operators are,
\begin{eqnarray}
\frac{\partial}{\partial t}Y_l(t;z)&=&U^{\dag}_{\rm original}(t;z_{1,\tau},z_{2,\tau})D(0;z)U_{\rm original}(t;z_{3,\tau},z_{4,\tau}),\nonumber\\
D(0;z)&=&i[H_{\rm sys}+\sum_{l}C_{l,\rm ext}S_{l,\rm c},Y_l]
     +\frac{1}{\sqrt{2}}S_{\rm sys}(z_{1,t}-iz_{2,t})Y_l-iY_{l}\frac{1}{\sqrt{2}}S_{\rm sys}(z_{3,t}+iz_{4,t}),\nonumber\\
\frac{\partial}{\partial t}O_{\rm ext}&=&i[\sum_{l}C_{l,\rm ext}S_{l,\rm c}+H_{\rm ext},O_{\rm ext}]\approx i[H_{\rm ext},O_{\rm ext}].\nonumber\\
     \label{apssle1}
\end{eqnarray}
The influence of the system on the external field is not considered here. Then, the dynamics of the external field is just the free dynamics,
\begin{eqnarray}
C_{l,\rm ext}(t)&=&e^{itH_{\rm ext}}C_{l,\rm ext}e^{-itH_{\rm ext}},\nonumber\\
\langle \psi_{\rm ext}|C_{l,\rm ext}(t)|\psi_{\rm ext}\rangle&=&C_l(t),\nonumber\\
\end{eqnarray}
with the initial state of the external field $|\psi_{\rm ext}\rangle$. Then we ``trace over'' the degrees of freedom of the external field in Eq.~(\ref{apssle1}) by taking the expectation value:
\begin{eqnarray}
\frac{\partial}{\partial t}Y_l(t;z)&=&\langle \psi_{\rm ext}|U^{\dag}_{\rm original}(t;z_{1,\tau},z_{2,\tau})D(0;z)U_{\rm original}(t;z_{3,\tau},z_{4,\tau})|\psi_{\rm ext}\rangle,\nonumber\\
&=&U^{\dag}_{\rm sys}(t;z_{1,\tau},z_{2,\tau})D'(0;z)U_{\rm sys}(t;z_{3,\tau},z_{4,\tau}),~{\rm where}\nonumber\\
D'(0;z)&=&i[H_{\rm sys}+\sum_{l}C_{l}(t)S_{l,\rm c},Y_l]
     +\frac{1}{\sqrt{2}}S_{\rm sys}(z_{1,t}-iz_{2,t})Y_l-iY_{l}\frac{1}{\sqrt{2}}S_{\rm sys}(z_{3,t}+iz_{4,t}).\nonumber\\
     \label{apssle2}
\end{eqnarray}
Here we have use the property $\langle\psi_{\rm ext}|\frac{\partial}{\partial t}Y_l(t;z)|\psi_{\rm ext}\rangle=\frac{\partial}{\partial t}Y_l(t;z)$ because the entanglement between the system and the external field is neglected. With Eq.~(\ref{apssle2}), we can follow the way in Sec.~\ref{secsle} and obtain the time-dependent stochastic equations.
\begin{eqnarray}
\frac{\partial}{\partial t}\mathcal{Y}(t,\xi,\eta)&=&\left(i\mathcal{H}+i\mathcal{C}(t)+i\frac{\xi_t}{\sqrt{2}}\mathcal{S}^c+\frac{\eta_t}{\sqrt{2}}\mathcal{S}^a\right)\mathcal{Y}(t,\xi,\eta),\nonumber\\
           \sum_m\mathcal{C}_{lm}(t)Y_{m}&=&\sum_{n}C_{n}(t)\left[S_{n},Y_l\right].\nonumber\\
\label{apssle4}
\end{eqnarray}
\end{widetext}


\begin{references}
%
%
%\bibitem{mb1}M. J. Hartmann, F. G. S. L. Brand$\rm\tilde{a}$o, and M. B. Plenio, Quantum %many-body phenomena in coupled cavity arrays, %\href{https://doi.org/10.1002/lpor.200810046}{Laser \& Photon. Rev. {\bf 2}, 527 (2008)}.
%
%
\bibitem{mb2} I. Bloch, J. Dalibard, and W. Zwerger, Many-body physics with ultracold gases, \href{https://doi.org/10.1103/RevModPhys.80.885}{Rev. Mod. Phys. {\bf 80}, 885 (2008)}.
%
%
\bibitem{mb3}M. A. Cazalilla and M. Rigol, Focus on dynamics and thermalization in isolated
quantum many-body systems, \href{https://doi.org/10.1088/1367-2630/12/5/055006}{New J. Phys. {\bf 12}, 055015 (2010)}.
%
%
\bibitem{mb4}M. Gring {\it et al.}, Relaxation and prethermalization in an isolated quantum system, \href{https://doi.org/10.1126/science.1224953}{Science 1224953 (2012)}.
%
%
\bibitem{mb5} T. Langen, R. Geiger, M. Kuhnert, B. Rauer and J. Schmiedmayer, Local emergence of thermal correlations in an isolated quantum many-body system, \href{https://doi.org/10.1038/NPHYS2739}{Nat. Phys.  {\bf 9}, 640 (2013)}.
%
%
\bibitem{mb6}M. Kuhnert, R. Geiger, T. Langen, M. Gring, B. Rauer, T. Kitagawa, E. Demler, D. Adu Smith, and J. Schmiedmayer, Multimode dynamics and emergence of a characteristic length scale in a one-dimensional quantum system, \href{https://doi.org/10.1103/PhysRevLett.110.090405}{Phys. Rev. Lett. {\bf 110}, 090405 (2013)}.
%
%
\bibitem{mb7}P. Jurcevic, B. P. Lanyon, P. Hauke, C. Hempel, P. Zoller, R. Blatt, and C. F. Roos
, Quasiparticle engineering and entanglement
propagation in a quantum many-body system, \href{https://doi.org/10.1038/nature13461}{Nature {\bf 511}, 202 (2014)}.
%
%
\bibitem{mb8}J. Eisert, M. Friesdorf, and C. Gogolin, Quantum many-body systems out of equilibrium, \href{https://doi.org/10.1038/NPHYS3215}{Nat. Phys.  {\bf 11}, 124 (2015)}.
%
%
\bibitem{mb9}A. M. Kaufman, M. E. Tai, A. Lukin, M. Rispoli, R. Schittko, P. M. Preiss, and Greiner, Quantum thermalization through entanglement in an isolated many-body system, \href{https://doi.org/10.1126/science.aaf6725}{Science {\bf 353}, 794 (2016)}.
%
%
\bibitem{mb10}B. Neyenhuis, J. Zhang, P. W. Hess, J. Smith, A. C. Lee, P. Richerme, Z.-X. Gong,
A. V. Gorshkov, and C. Monroe, Observation of prethermalization in long-range
interacting spin chains, \href{https://doi.org/10.1126/sciadv.1700672}{Sci. Adv. {\bf 3}, e1700672 (2017)}.
%
%
\bibitem{qtd1}M. Campisi, P. Talkner, and P. H$\rm\ddot{a}$nggi, Fluctuation theorem for arbitrary open quantum systems, \href{https://doi.org/10.1103/PhysRevLett.102.210401}{Phys. Rev. Lett. {\bf 102}, 210401 (2009)}.
%
%
%\bibitem{qtd2}M. Campisi, D. Zueco. and P. Talkner, Thermodynamic anomalies in open quantum systems: Strong coupling effects in the isotropic XY model, \href{https://doi.org/10.1016/j.chemphys.2010.04.026}{J. Chemphys. {\bf 375}, 187 (2010)}.
%
%
\bibitem{qtd3}M. Esposito, M. A. Ochoa, and M. Galperin, Nature of heat in strongly coupled open quantum systems, \href{https://doi.org/10.1103/PhysRevB.92.235440}{Phys. Rev. B {\bf 92}, 235440 (2015)}.
%
%
\bibitem{qtd4}M. Esposito, M. A. Ochoa, and M. Galperin, Quantum thermodynamics: a nonequilibrium Green's function approach, \href{https://doi.org/10.1103/PhysRevLett.114.080602}{Phys. Rev. Lett. {\bf 114}, 080602 (2015)}.
%
%
\bibitem{qtd5}U. Seifert, First and second law of thermodynamics at strong coupling, \href{https://doi.org/10.1103/PhysRevLett.116.020601}{Phys. Rev. Lett. {\bf 116}, 020601 (2016)}.
%
%
\bibitem{qtd6}J. Liu, H. Xu, B Li, and C. Wu, Energy transfer in the nonequilibrium spin-boson model: from weak to strong coupling, \href{https://doi.org/10.1103/PhysRevE.96.012135}{Phys. Rev. E {\bf 96}, 012135 (2017)}.
%
%
\bibitem{qtd7}C. Jarzynski, Stochastic and macroscopic thermodynamics of strongly coupled systems, \href{https://doi.org/10.1103/PhysRevX.7.011008}{Phys. Rev. X {\bf 7}, 011008 (2017)}.
%
%
\bibitem{qtd8}A. Bruch, C. Lewenkopf, and F. V. Oppen, Landauer-B$\rm\ddot{u}$ttiker approach to strongly coupled quantum thermodynamics: inside-outside duality of entropy evolution, \href{https://doi.org/10.1103/PhysRevLett.120.107701}{Phys. Rev. Lett. {\bf 120}, 107701 (2018)}.
%
%
\bibitem{qtd9}M. Perarnau-Llobet, H. Wilming, A. Riera, R. Gallego, and J. Eisert, Strong coupling corrections in quantum thermodynamics, \href{https://doi.org/10.1103/PhysRevLett.120.120602}{Phys. Rev. Lett. {\bf 120}, 120602 (2018)}.
%
%
%\bibitem{qtd10}J.-T. Hsiang and B.-L. Hu, Quantum thermodynamics at strong coupling:
%operator thermodynamic functions and relations, \href{https://doi.org/10.3390/e20060423}{Entropy %{\bf 20}, 423 (2018)}.
%
%
\bibitem{qbio1}N. Lambert, Y.N. Chen, Y.C. Chen, C.M. Li, G.Y. Chen, F. Nori, Quantum biology, \href{https://doi.org/10.1038/nphys2474}{Nat. Phys. {\bf 9}, 10-18 (2013)}.
%
%
\bibitem{qbio2}L.G. Mourokh, F. Nori, Energy transfer efficiency in the chromophore network strongly coupled to a vibrational mode, \href{https://doi.org/10.1103/PhysRevE.92.052720}{Phys. Rev. E {\bf 92}, 052720 (2015)}.
%
%
\bibitem{qbio3}H.B. Chen, N. Lambert, Y.C. Cheng, Y.N. Chen, F. Nori, Using non-Markovian measures to evaluate quantum master equations for photosynthesis, \href{https://doi.org/10.1038/srep12753}{Sci. Rep. {\bf 5}, 12753 (2015).}
%
%
\bibitem{qbio4}B.X. Wang, M.J. Tao, Q. Ai, T. Xin, N. Lambert, D. Ruan, Y.C. Cheng, F. Nori, F.G. Deng, G.L. Long, Efficient quantum simulation of photosynthetic light harvesting, \href{https://doi.org/10.1038/s41534-018-0102-2}{npj Quantum Inf. {\bf 4}, 52 (2018)}.
%
%
\bibitem{photosyn1}T. Brixner, J. Stenger, H. M. Vaswani, M. Cho, R. E. Blankenship, and G. R. Fleming, Two-dimensional spectroscopy of electronic couplings in
photosynthesis, \href{https://doi.org/10.1038/nature03429}{Nature {\bf 434}, 625 (2005)}.
%
%
\bibitem{photosyn2}A. W. Chin, J. Prior, R. Rosenbach, F. Caycedo-Soler, S. F. Huelga, and M. B. Plenio, The role of non-equilibrium vibrational structures in electronic coherence and recoherence in pigment–protein complexes, \href{https://doi.org/10.1038/nphys2515}{Nat. Phys. {\bf 9}, 113 (2013)}.

%
%
\bibitem{photosyn3}A. Halpin, P. J. M. Johnson, R. Tempelaar, R. S. Murphy, J. Knoester, T. L. C. Jansen, and R. J. D. Miller, Coherent exciton-vibrational dynamics and energy
transfer in conjugated organics, \href{https://doi.org/10.1038/NCHEM.1834}{Nat. Chem. {\bf 6}, 196 (2014)}.
%
%
\bibitem{photosyn4}T. R. Nelson, D. Ondarse-Alvarez, N. Oldani, B. Rodriguez-Hernandez, L. Alfonso-Hernandez, J. F. Galindo,
V. D. Kleiman, S. Fernandez-Alberti, A. E. Roitberg, and S. Tretiak, Two-dimensional spectroscopy of a molecular
dimer unveils the effects of vibronic coupling
on exciton coherences, \href{https://doi.org/10.1038/s41467-018-04694-8}{Nat. Commun. {\bf 9}, 2316 (2018)}.
%
%
\bibitem{photosyn5}M. Maiuri, E. E. Ostroumov, R. G. Saer, R. E. Blankenship, and G. D. Scholes, Coherent wavepackets in the Fenna-Matthews-Olson complex are robust to excitonic-structure
perturbations caused by mutagenesis, \href{https://doi.org/10.1038/NCHEM.2910}{Nat. Chem. {\bf 10}, 177 (2018)}.
%
%
\bibitem{drib1}J. Reichert, P. Nalbach, and M. Thorwart, Dynamics of a quantum two-state system in a linearly driven quantum bath, \href{https://doi.org/10.1103/PhysRevA.94.032127}{Phys. Rev. A {\bf 94}, 032127 (2016)}.
%
%
\bibitem{drib2}H. Grabert, P. Nalbach, J. Reichert, and M. Thorwart, Nonequilibrium Response of Nanosystems Coupled to Driven Quantum Baths, \href{https://doi.org/10.1021/acs.jpclett.6b00703}{J. Phys. Chem. Lett. {\bf 7}, 2015 ( 2016)}.
%
%
\bibitem{drib3}H. Grabert and M. Thorwart, Quantum mechanical response to a driven Caldeira-Leggett bath, \href{https://doi.org/10.1103/PhysRevE.98.012122}{Phys. Rev. E {\bf 98}, 012122 (2018)}.
%
%
\bibitem{mebi1}S Gasparinetti, P Solinas, A Braggio, and M Sassetti, Heat-exchange statistics in driven open quantum systems, \href{https://doi.org/10.1088/1367-2630/16/11/115001}{New J. Phys. {\bf 16}, 115001 (2014)}.
%
%
\bibitem{mebi2}M. Carrega, P. Solinas, M. Sassetti, and U. Weiss, Energy exchange in driven open quantum systems at strong coupling, \href{https://doi.org/10.1103/PhysRevLett.116.240403}{Phys. Rev. Lett. {\bf 116}, 240403 (2016)}.
%
%
\bibitem{mebi3}L. Song and Q. Shi, Hierarchical equations of motion method applied to nonequilibrium heat transport in model molecular junctions: transient heat current and high-order moments of the current operator, \href{https://doi.org/10.1103/PhysRevB.95.064308}{Phys. Rev. B {\bf 95}, 064308 (2017)}.
%
%
\bibitem{qle1}G. W. Ford, J. T. Lewis, and R. F. O'Connell, Quantum Langevin equation, \href{https://doi.org/10.1103/PhysRevA.37.4419}{Phys. Rev. A {\bf 37}, 4410 (1988)}.
%
%
\bibitem{qle2}C.-P. Sun and L.-H. Yu, Exact dynamics of a quantum dissipative system in a constant external field, \href{https://doi.org/10.1103/PhysRevA.51.1845}{Phys. Rev. A {\bf 51}, 1845 (1995)}.
%
%
\bibitem{qle3}J. J. Hope, Theory of input and output of atoms from an atomic trap, \href{https://doi.org/10.1103/PhysRevA.55.R2531}{Phys. Rev. A {\bf 55}, R2531 (1997)}.
%
%
\bibitem{qle4}P. Wang, A. M. Tartakovsky, and D. M. Tartakovsky, Probability density function method for Langevin equations with colored noise, \href{https://doi.org/10.1103/PhysRevLett.110.140602}{Phys. Rev. Lett. {\bf 110}, 140602 (2013)}.
%
%
\bibitem{qle5}T. Brett and T. Galla, Stochastic processes with distributed delays: chemical Langevin wquation and linear-noise approximation, \href{https://doi.org/10.1103/PhysRevLett.110.250601}{Phys. Rev. Lett. {\bf 110}, 250601 (2013)}.
%
%
\bibitem{qle6}L. Stella, C. D. Lorenz, and L. Kantorovich, Generalized Langevin equation: An efficient approach to nonequilibrium molecular dynamics of open systems, \href{https://doi.org/10.1103/PhysRevB.89.134303}{Phys. Rev. B {\bf 89}, 134303 (2014)}.
%
%
%\bibitem{qle7}J.-T. Hsiang and B.-L. Hu, Nonequilibrium steady state in open quantum
%systems: Influence action, stochastic equation and power balance, %\href{https://doi.org/10.1016/j.aop.2015.07.009}{Ann. Phys. {\bf 362}, 139 (2015)}.
%
%
\bibitem{qleb1}J. Rosa and M. W. Beims, Dissipation and transport dynamics in a ratchet coupled to a discrete bath, \href{https://doi.org/10.1103/PhysRevE.78.031126}{Phys. Rev. E {\bf 78}, 031126 (2008)}.
%
%
\bibitem{qleb2}H. Hasegawa, Specific heat anomalies of small quantum systems subjected to finite baths, \href{https://doi.org/10.1063/1.3669485}{J. Math. Phys. {\bf 52}, 123301 (2011)}.
%
%
\bibitem{qleb3}H. Hasegawa, Classical small systems coupled to finite baths, \href{10.1103/PhysRevE.83.021104}{Phys. Rev. E {\bf 83}, 021104 (2011)}.
%
%
%\bibitem{acem1}I. R. Senitzky, Induced and spontaneous emission in a coherent field, \href{https://doi.org/10.1103/PhysRev.111.3}{Phys. Rev. {\bf 111}, 3 (1958)}.
%
%
%\bibitem{acem2}I. R. Senitzky, Dissipation in quantum mechanics. The two-level system, \href{https://doi.org/10.1103/PhysRev.131.2827}{Phys. Rev. {\bf 131}, 2827 (1963)}.
%
%
%\bibitem{acem3}G. S. Agarwal, Master-equation approach to spontaneous emission, \href{https://doi.org/10.1103/PhysRevA.2.2038}{Phys. Rev. A {\bf 2}, 2038 (1970)}.
%
%
%\bibitem{deco1}L. Viola and S. Lloyd, Dynamical suppression of decoherence in two-state quantum systems, \href{https://doi.org/10.1103/PhysRevA.58.2733}{Phys. Rev. A {\bf 58}, 2733 (1998)}.
%
%
%\bibitem{deco2}T. Yu and J. H. Eberly, Phonon decoherence of quantum entanglement: Robust and fragile states, \href{https://doi.org/10.1103/PhysRevB.66.193306}{Phys. Rev. B {\bf 66}, 193306 (2002)}.
%
%
%\bibitem{deco3}T. Yu and J. H. Eberly, Qubit disentanglement and decoherence via dephasing, \href{https://doi.org/10.1103/PhysRevB.68.165322}{Phys. Rev. B {\bf 68}, 165322 (2003)}.
%
%
%\bibitem{deco4}A. R. R. Carvalho, F. Mintert, and A. Buchleitner, Decoherence and Multipartite Entanglement, \href{https://doi.org/10.1103/PhysRevLett.93.230501}{Phys. Rev. Lett. {\bf 93}, 230501 (2004)}.
%
%
%\bibitem{Jing}J. Jing, L.-A. Wu, M. Byrd, J. Q. You, T. Yu, and Z.-M. Wang, Nonperturbative Leakage Elimination Operators and Control of a Three-Level System, \href{https://journals.aps.org/prl/abstract/10.1103/PhysRevLett.114.190502}{Phys. Rev. Lett. {\bf 114}, 190502 (2015)}.
%
%
\bibitem{Diosi} L. Di\'{o}si, N. Gisin, and W. T. Strunz, Non-Markovian quantum state diffusion, \href{https://doi.org/10.1103/PhysRevA.58.1699}{Phys. Rev. A {\bf 58}, 1699 (1998)}.
%
%
\bibitem{qsd1}T. Yu, L. Di$\rm\acute{o}$si, N. Gisin, and W. T. Strunz, Non-Markovian quantum-state diffusion: Perturbation approach, \href{https://doi.org/10.1103/PhysRevA.60.91}{Phys. Rev. A {\bf 60}, 91 (1999)}.
%
%
\bibitem{qsd2} J. Jing, X. Zhao, J. Q. You,  and T. Yu, Time-local quantum-state-diffusion equation for multilevel quantum systems,\href{https://doi.org/10.1103/PhysRevA.85.042106}{Phys. Rev. A {\bf 85}, 042106 (2012)}.
%
%
\bibitem{qsd3} Z.-Z. Li, C.-T. Yip, H.-Y. Deng, M. Chen, T. Yu, J. Q. You, and C.-H. Lam, Approach to solving spin-boson dynamics via non-Markovian quantum trajectories, \href{https://doi.org/10.1103/PhysRevA.90.022122}{Phys. Rev. A {\bf 90}, 022122 (2014)}.
%
%
\bibitem{qsd4} D.-W. Luo, C.-H. Lam, L.-A. Wu, T. Yu, H.-Q. Lin, and J. Q. You, Higher-order solutions to non-Markovian quantum dynamics via a hierarchical functional derivative, \href{https://doi.org/10.1103/PhysRevA.92.022119}{Phys. Rev. A {\bf 92}, 022119 (2015)}.
%
%
\bibitem{polatran1}A. W$\rm\ddot{u}$rger, Strong-coupling theory for the spin-phonon model, \href{https://doi.org/10.1103/PhysRevB.57.347}{Phys. Rev. B {\bf 57}, 347 (1998)}.
%
%
\bibitem{polatran2}I. Wilson-Rae and A. Imamo$\rm \breve{g}$lu, Quantum dot cavity-QED in the presence of strong electron-phonon interactions, \href{https://doi.org/10.1103/PhysRevB.65.235311}{Phys. Rev. B {\bf 65}, 235311 (2002)}.
%
%
\bibitem{polatran3}D. P. S. McCutcheon, N. S. Dattani, E. M. Gauger, B. W. Lovett, and A. Nazir, A general approach to quantum dynamics using a variational master equation: Application to phonon-damped Rabi rotations in quantum dots, \href{https://doi.org/10.1103/PhysRevB.84.081305}{Phys. Rev. B {\bf 84}, 081305(R) (2011)}.
%
%
\bibitem{polatran4}C. Zhao, Z. L$\rm\ddot{u}$, and H. Zheng, Entanglement evolution and quantum phase transition of biased s = 1/2 spin-boson model, \href{https://doi.org/10.1103/PhysRevE.84.011114}{Phys. Rev. E {\bf 84}, 011114 (2011)}.
%
%
\bibitem{polatran5}C. Roy and S. Hughes, Phonon-dressed Mollow triplet in the regime of cavity quantum electrodynamics: excitation-induced dephasing and nonperturbative cavity feeding effects, \href{https://doi.org/10.1103/PhysRevLett.106.247403}{Phys. Rev. Lett. {\bf 106}, 247403 (2011)}.
%
%
\bibitem{polatran6}Z. L$\rm\ddot{u}$ and H. Zheng, Effects of counter-rotating interaction on driven tunneling dynamics: coherent destruction of tunneling and Bloch-Siegert shift, \href{https://doi.org/10.1103/PhysRevA.86.023831}{Phys. Rev. A {\bf 86}, 023831 (2012)}.
%
%
\bibitem{polatran7}Y. Yan, Z. L$\rm\ddot{u}$, and H. Zheng, Effects of counter-rotating-wave terms of the driving field on the spectrum of resonance fluorescence, \href{https://doi.org/10.1103/PhysRevA.88.053821}{Phys. Rev. A {\bf 88}, 053821 (2013)}.
%
%
\bibitem{polatran8}R. Manson, K. Roy-Choudhury, and S. Hughes, Polaron master equation theory of pulse-driven phonon-assisted population inversion and single-photon emission from
quantum-dot excitons, \href{https://doi.org/10.1103/PhysRevB.93.155423}{Phys. Rev. B {\bf 93}, 155423 (2016)}.
%
%
\bibitem{polatran9}C. Gustin and S. Hughes, Pulsed excitation dynamics in quantum-dot–cavity systems: Limits to optimizing the fidelity of on-demand single-photon sources, \href{https://doi.org/10.1103/PhysRevB.98.045309}{Phys. Rev. B {\bf 98}, 045309 (2018)}.
%
%
\bibitem{hie1}V. E. Shapiro and V. M. Loginov, "Formulae of differentiation" and their use for solving stochastic equations, \href{https://doi.org/10.1016/0378-4371(78)90198-X}{Physica A {\bf 91}, 563 (1978)}.
%
%
\bibitem{hie2}Y. Tanimura and R. Kubo, Time evolution of a quantum system in contact with a nearly Gaussian-Markoffian noise bath, \href{https://doi.org/10.1143/JPSJ.58.101}{J. Phys. Soc. Jpn. {\bf 58}, 101 (1989)}.
%
%
\bibitem{hie3}Y. Yan, F. Yang, Y. Liu, J. Shao, Hierarchical approach based on stochastic decoupling to dissipative systems, \href{https://doi.org/10.1016/j.cplett.2004.07.036}{Chem. Phys. Lett. {\bf 395}, 216 (2004)}.
%
%
\bibitem{hie4}R. Xu, P. Cui, X. Li, Y. Mo, and Y. Yan, Exact quantum master equation via the calculus on path integrals, \href{https://doi.org/10.1063/1.1850899}{J. Chem. Phys. {\bf 122}, 041103 (2005)}.
%
%
\bibitem{hie5}A. Ishizaki, Y. Tanimura, Quantum dynamics of system strongly coupled to low-temperature colored noise bath: reduced hierarchy equations approach, \href{https://doi.org/10.1143/jpsj.74.3131}{J. Phys. Soc. Jpn. {\bf 74}, 3131 (2005)}.
%
\bibitem{hie5.5}X. Yin, J. Ma, X.G. Wang, F. Nori, Spin squeezing under non-Markovian channels by the hierarchy equation method, \href{https://doi.org/10.1103/PhysRevA.86.012308}{Phys. Rev. A {\bf 86}, 012308 (2012)}.
%
\bibitem{hie6}C. Duan, Z. Tang, J. Cao, and J. Wu, Zero-temperature localization in a sub-Ohmic spin-boson model investigated by an extended hierarchy equation of motion, \href{https://doi.org/10.1103/PhysRevB.95.214308}{Phys. Rev. B {\bf 95}, 214308 (2017)}.
%
%
\bibitem{hie7}C.-Y. Hsieh and J. Cao, A unified stochastic formulation of dissipative quantum dynamics. I. Generalized hierarchical equations, \href{https://doi.org/10.1063/1.5018725}{J. Chem. Phys. {\bf 148}, 014103 (2018)}.
%
%
\bibitem{pathi} W.-M. Zhang, P.-Y. Lo, H.-N. Xiong, M. W.-Y. Tu, and F. Nori, General non-Markovian dynamics of open quantum systems, \href{https://doi.org/10.1103/PhysRevLett.109.170402}{Phys. Rev. Lett. {\bf 109}, 170402 (2012)}.
%
%
\bibitem{sfd0}Y. Tanimura, Stochastic Liouville, Langevin, Fokker–Planck, and Master Equation Approaches to Quantum Dissipative Systems, \href{https://doi.org/10.1143/JPSJ.75.082001}{J. Phys. Soc. Jpn. {\bf 75}, 082001 (2006)}.
%
%
\bibitem{sfd1} J. T. Stockburger and C. H. Mak, Dynamical simulation of current fluctuations in a dissipative two-state system, \href{https://doi.org/10.1103/PhysRevLett.80.2657}{Phys. Rev. Lett. {\bf 80}, 2657 (1997)}.
%
%
\bibitem{sfd2} J. T. Stockburger and H. Grabert, Exact {\it c}-number Representation of Non-Markovian Quantum Dissipation, \href{https://doi.org/10.1103/PhysRevLett.88.170407}{Phys. Rev. Lett. {\bf 88}, 170407 (2002)}.
%
%
\bibitem{sfd2.5}J. Shao, Decoupling quantum dissipation interaction via stochastic fields, \href{https://doi.org/10.1063/1.1647528}{J. Chem. Phys. {\bf 120}, 5053 (2004)}.
%
%
\bibitem{sfd3} W. Koch and F. Gro$\rm{\ss}$mann, Non-Markovian dissipative semiclassical dynamics, \href{https://doi.org/10.1103/PhysRevLett.100.230402}{Phys. Rev. Lett. {\bf 100}, 230402 (2008)}.
%
%
\bibitem{sfd4} Y.-A. Yan and Y. Zhou, Hermitian non-Markovian stochastic master equations for quantum dissipative dynamics, \href{https://doi.org/10.1103/PhysRevA.92.022121}{Phys. Rev. A {\bf 92}, 022121 (2015)}.
%
%
\bibitem{sfd5} Y.-A. Yan and J. Shao, Stochastic description of quantum Brownian dynamics, \href{https://doi.org/10.1007/s11467-016-0570-9}{Front. Phys. 11, 110309 (2016)}.
%
%
\bibitem{idcoupling1}S.-B. Zheng and G.-C. Guo, Efficient scheme for two-atom entanglement and quantum information processing in cavity QED, \href{https://doi.org/10.1103/PhysRevLett.85.2392}{Phys. Rev. Lett. {\bf 85}, 2392 (2000)}.
%
%
\bibitem{idcoupling2}A. F. van Loo, A. Fedorov, K. Lalumi${\rm \grave{e}}$re, B. C. Sanders, A. Blais, and A. Wallraff, Photon-mediated interactions between distant artificial atoms, \href{https://doi.org/10.1126/science.1244324}{Science {\bf 342}, 1494 (2013)}.
%
%
\bibitem{idcoupling3}C. Song, K. Xu, W. Liu, C.-P. Yang, S.-B. Zheng, H. Deng, Q. Xie, K. Huang, Q. Guo, L. Zhang, P. Zhang, D. Xu, D. Zheng, X. Zhu, H. Wang, Y.-A. Chen, C.-Y. Lu, S. Han, and J.-w. Pan, 10-qubit entanglement and parallel logic operations with a superconducting circuit, \href{https://doi.org/10.1103/PhysRevLett.119.180511}{Phys. Rev. Lett. {\bf 119}, 180511 (2017)}.
%
%
\bibitem{idcoupling4}K. Xu, J.-J. Chen, Y. Zeng, Y.-R. Zhang, C. Song, W. Liu, Q. Guo, P. Zhang, D. Xu, H. Deng, K. Huang, H. Wang, X. Zhu, D. Zheng, and H. Fan, Emulating many-body localization with a superconducting quantum processor, \href{https://doi.org/10.1103/PhysRevLett.120.050507}{Phys. Rev. Lett. {\bf 120}, 050507 (2018)}.
%
%
\bibitem{oqs10}C.-Y. Hsieh and J. Cao, A unified stochastic formulation of dissipative quantum dynamics. II. Beyond linear response of spin baths, \href{https://doi.org/10.1063/1.5018726}{J. Chem. Phys. {\bf 148}, 014104 (2018)}.
%
%
%\bibitem{sct1}J. Q. You and F. Nori, Quantum information processing with superconducting qubits in a microwave field, \href{https://doi.org/10.1103/PhysRevB.68.064509}{Phys. Rev. B {\bf 68}, 064509 (2003)}.
%
%
%\bibitem{sct2}J. Q. You and F. Nori, Superconducting Circuits and Quantum Information, \href{https://doi.org/10.1063/1.2155757}{Phys. Today {\bf 58}, 42 (2005)}.
%
%
%\bibitem{sct3}W. Qin, A. Miranowicz, P.-B. Li, X.-Y. L\"{u}, J. Q. You, and F. Nori, Exponentially enhanced light-matter interaction, cooperativities, and steady-state
%entanglement using parametric amplification, \href{https://doi.org/10.1103/PhysRevLett.120.093601}{Phys. Rev. Lett. {\bf 120}, 093601 (2018)}.
%
%
%\bibitem{sc1}A. Wallraff et al., Strong coupling of a single photon to a superconducting qubit using circuit quantum electrodynamics, \href{https://doi.org/10.1038/nature02851}{Nature {\bf 431}, 162 (2004)}.
%
%
%\bibitem{sc2}J. P. Reithmaier et al., Strong coupling in a single quantum dot semiconductor %microcavity system, \href{https://doi.org/10.1038/nature02969}{Nature {\bf 432}, 197 (2004)}.
%

%
%\bibitem{sc3}T. Aoki {\it et al.}, Observation of strong coupling between one atom
%and a monolithic microresonator, \href{https://doi.org/10.1038/nature05147}{Nature {\bf 443}, 671 (2006)}.
%
%
%\bibitem{sc4}Y. Colombe, T. Steinmetz, G. Dubois, F. Linke, D. Hunger, and J. Reichel, Strong atom-field coupling for Bose-Einstein condensates in an optical cavity on a chip, \href{https://doi.org/10.1038/nature06331}{Nature {\bf 450}, 272 (2007)}.
%
%
%\bibitem{sc5}G. A. Steele {\it et al.}, Strong coupling between single-electron tunneling and
%nanomechanical motion, \href{https://doi.org/10.1126/science.1176076}{Science {\bf 325}, 1103 (2009)}.
%
%
%\bibitem{sc6}S. Gr$\rm \ddot{o}$blacher, K. Hammerer, M. R. Vanner, and M. Aspelmeyer, Observation of strong coupling between a micromechanical resonator and an optical cavity field, \href{https://doi.org/10.1038/nature08171}{Nature {\bf 460}, 724 (2009)}.
%
%
%\bibitem{sc7}J. D. Teufel, D. Li, M. S. Allman, K. Cicak, A. J. Sirois, J. D. Whittaker, and R. W. Simmonds, Circuit cavity electromechanics in the strong-coupling regime, \href{https://doi.org/10.1038/nature09898}{Nat. Phys.  {\bf 6}, 772 (2010)}.
%


%
%\bibitem{usc1}T. Niemczyk {\it et al.}, Circuit quantum electrodynamics in the %ultrastrong-coupling regime, \href{https://doi.org/10.1038/NPHYS1730}{Nature {\bf 460}, 724 %(2009)}.
%

%
%\bibitem{usc2}P. Forn-D\'{\i}az et al., Observation of the Bloch-Siegert shift in a qubit-oscillator System in the ultrastrong coupling regime, \href{https://doi.org/10.1103/PhysRevLett.105.237001}{Phys. Rev. Lett. {\bf 105}, 237001 (2010)}.
%
%
%\bibitem{usc3}G. Scalari {\it et al.}, Ultrastrong coupling of the cyclotron transition of a 2D electron gas to a THz Metamaterial, \href{https://doi.org/10.1126/science.1216022}{Science {\bf 335}, 1323 (2012)}.
%
%
%\bibitem{usc4}Z. Chen {\it et al.}, Single-photon-driven high-order sideband transitions in an ultrastrongly coupled circuit-quantum-electrodynamics system, \href{https://doi.org/10.1103/PhysRevA.96.012325}{Phys. Rev. A {\bf 96}, 012325 (2017)}.
%

%
%\bibitem{usc5}L. Garziano, R. Stassi, V. Macr\`{\i}, A. F. Kockum, S. Savasta, and F. Nori, Multiphoton quantum Rabi oscillations in ultrastrong cavity QED, \href{https://doi.org/10.1103/PhysRevA.92.063830}{Phys. Rev. A {\bf 92}, 063830 (2015)}.
%
%
%\bibitem{usc6}M. Cirio, Kamanasish, N. Lambert, and F. Nori, Amplified Optomechanical transduction of virtual radiation pressure, \href{https://doi.org/10.1103/PhysRevLett.119.053601}{Phys. Rev. Lett. {\bf 119}, 053601 (2017)}.
%
%
%\bibitem{usc7}C. S. Mu\~{n}oz, F. Nori, and S. D. Liberato, Resolution of superluminal signalling in non-perturbative cavity quantum electrodynamics, \href{https://doi.org/10.1038/s41467-018-04339-w}{Nat. comm. {\bf 9}, 1924 (2018)}.
%




%
\bibitem{hstrans}J. Hubbard, Calculation of partition functions, \href{10.1103/10.1103/PhysRevLett.3.77}{Phys. Rev. Lett. {\bf 3}, 77 (1959)}.
%
%
\bibitem{osbook1}C. W. Gardiner and P. Zoller, Quantum Noise (Springer, 2004).
%
%
\bibitem{c-l model} A. J. Leggett, S. Chakravarty, A. T. Dorsey, M. P. A. Fisher, A. Garg and W. Zwerger, Dynamics of the dissipative two-state system, \href{https://doi.org/10.1103/RevModPhys.59.1}{Rev. Mod. Phys. {\bf 59}, 1 (1987)}.
%
%
\bibitem{mge} W. Magnus, On the exponential solution of differential equations for a linear operator, \href{https://doi.org/10.1002/cpa.3160070404}{Comm. Pure and Appl. Math. {\bf 7}, 649-673 (1954)}.
%
%
\bibitem{mge2}S. Blanes, F. Casas, J. A. Oteo, and J. Ros, The Magnus expansion and some of its applications, \href{https://doi.org/10.1016/j.physrep.2008.11.001}{Phys. Rep. {\bf 470}, 151-238 (2009)}.
%
%
\bibitem{ods}A. Fruchtman, N. Lamber, and E. Gauger, When do perturbative approaches accurately capture the dynamics of complex quantum systems? \href{https://doi.org/10.1038/srep28204}{Sci. Rep. {\bf 6}, 28204 (2016)}.
%
%
\bibitem{fft}M. Frigo and S. G. Johnson, The Design and Implementation of FFTW3, \href{https://doi.org/10.1109/JPROC.2004.840301}{Proc. IEEE {\bf 93}, 216-231 (2005)}.
%
%
\bibitem{mtt}S. N. Shevchenko, A. I. Ryzhov, and F. Nori, Low-frequency spectroscopy for quantum multi-level systems, \href{https://doi.org/10.1103/PhysRevB.98.195434}{Phys. Rev. B {\bf98}, 195434 (2018)}.
%
%
\bibitem{uosct5}W. Qin, A. Miranowicz, P.-B. Li, X.-Y. L\"{u}, J. Q. You, and F. Nori, Exponentially enhanced light-matter interaction, cooperativities, and steady-state
entanglement using parametric amplification, \href{https://doi.org/10.1103/PhysRevLett.120.093601}{Phys. Rev. Lett. {\bf 120}, 093601 (2018)}.
%
%
\bibitem{uosct2}C. Leroux, L. C. G. Govia, and A. A. Clerk, Enhancing Cavity Quantum Electrodynamics via Antisqueezing: Synthetic Ultrastrong Coupling, \href{https://doi.org/10.1103/PhysRevLett.120.093602}{Phys. Rev. Lett. {\bf 120}, 093602 (2018)}.
%
%
\bibitem{qdots1}L. Besombes, K. Kheng, L. Marsal, and H. Mariette, Acoustic phonon broadening mechanism in single quantum dot emission, \href{https://doi.org/10.1103/PhysRevB.63.155307}{Phys. Rev. B {\bf 63}, 155307 (2001)}.
%
%
\bibitem{qdots2}S. Weiler, A. Ulhaq, S. M. Ulrich, D. Richter, M. Jetter, P. Michler, C. Roy, and S. Hughes, Phonon-assisted incoherent excitation of a quantum dot and its emission properties, \href{https://doi.org/10.1103/PhysRevB.86.241304}{Phys. Rev. B {\bf 86}, 241304(R) (2012)}.
%
%
\bibitem{papi1}S. Hughes and H. J. Carmichael, Phonon-mediated population inversion in a semiconductor quantum-dot cavity system, \href{https://doi.org/10.1088/1367-2630/15/5/053039}{New J. Phys. {\bf 15}, 053039 (2013)}.
%
%
\bibitem{papi2}J. H. Quilter, A. J. Brash, F. Liu, M. Gl$\rm \ddot{a}$ssl, A. M. Barth, V. M. Axt, A. J. Ramsay, M. S. Skolnick, and A. M. Fox, Phonon-assisted population inversion of a single InGaAs/GaAs quantum dot by pulsed laser excitation,  \href{https://doi.org/10.1103/PhysRevLett.114.137401}{Phys. Rev. Lett. {\bf 114}, 137401 (2014)}.
%
%
\bibitem{papi3}P.-L. Ardelt, L. Hanschke, K. A. Fischer, K. M$\rm \ddot{u}$ller, A. Kleinkauf, M. Koller, A. Bechtold, T. Simmet, J. Wierzbowski, H. Riedl, G. Abstreiter, and J. J. Finley, Dissipative preparation of the exciton and biexciton in self-assembled quantum dots on picosecond time scales, \href{https://doi.org/10.1103/PhysRevB.90.241404}{Phys. Rev. B {\bf 90}, 241404(R) (2014)}.
%
\end{references}
\end{document}